\documentclass[preprint2]{aastex6} 

\slugcomment{To be submitted to Astrophys. J.}

\shorttitle{Long-Lived Quasi-Periodic Pulsations}
\shortauthors{Dennis et al.}

\begin{document}

\title{Detection and Interpretation of Long-Lived X-ray Quasi-Periodic Pulsations in the X-class Solar Flare on 2013 May 14}

\author{Brian R. Dennis\altaffilmark{1}, Anne K. Tolbert\altaffilmark{2},
Andrew Inglis\altaffilmark{2}, Jack Ireland\altaffilmark{3}, Tongjiang Wang\altaffilmark{2}, and Gordon D. Holman}
\affil{Solar Physics Laboratory, Code 671, Heliophysics Science Division, NASA Goddard Space Flight Center, Greenbelt, MD 20771, USA}

\altaffiltext{1}{brian.r.dennis@nasa.gov}

\and

\author{Laura A. Hayes and Peter T. Gallagher}
\affil{School of Physics, Trinity College Dublin, Dublin 2, Ireland}


\altaffiltext{2}{The Catholic University of America, 620 Michigan Ave.~NE, Washington, DC 20064, USA}
\altaffiltext{3}{ADNET Systems, Inc.\ at NASA Goddard Space Flight Center, Greenbelt, MD 20771, USA}

\date{\today}

\begin{abstract}

Quasi-periodic pulsations (QPP) seen in the time derivative of the GOES soft X-ray light curves are analyzed for the \textsf{near-limb} X3.2 event on 14 May 2013. The pulsations are apparent for a total of at least two hours from the impulsive phase to well into the decay phase, with a total of 163 distinct pulses evident to the naked eye. A wavelet analysis shows that the characteristic time scale of these pulsations increases systematically from $\sim$25 s at 01:10 UT, the time of the GOES peak, to $\sim$100 s at 02:00 UT. A second `ridge' in the wavelet power spectrum, {most likely associated with flaring emission from a different active region,} shows an increase from $\sim$40 s at 01:40 UT to $\sim$100 s at 03:10 UT. We assume that the QPP {that produced the first ridge} result from vertical kink-mode oscillations of the newly formed loops following magnetic reconnection in the coronal current sheet. This allows us to estimate the magnetic field strength as a function of altitude given the density, loop length, and QPP time scale as functions of time determined from the GOES light curves and RHESSI images. The calculated magnetic field strength of the newly formed loops ranges from $\sim$500 G at an altitude of 24 Mm to {a low value of $\sim$10 G} at 60 Mm, in general agreement with the expected values at these altitudes. Fast sausage mode oscillations are also discussed and cannot be ruled out as an alternate mechanism for producing the QPP.

\end{abstract}

\keywords{Sun: flares --- Sun: corona --- Sun: X-rays, gamma rays --- Sun: oscillations}


\section{INTRODUCTION}
\label{intro}

The impulsive phase of a solar flare is generally characterized by multiple peaks in the hard X-ray (HXR) and microwave light curves.  This behavior was first noticed in HXRs and microwaves by \cite{1969ApJ...155L.117P} and in microwaves by \cite{1969ApJ...158L.127J}.  \cite{1968ApJ...153L..59N} had noted that the soft X-ray (SXR) light curve of many flares closely matched the time integral of the microwave light curve and this became the basis for what is now known as the Neupert Effect \citep{1968ApJ...153L..59N, 1991BAAS...23R1064H, 2005ApJ...621..482V}. \cite{1993SoPh..146..177D} showed that this general relationship is more clearly evident if the time derivative of the SXR light curve is used since it tends to match the HXR light curve through the impulsive phase \citep[see also ][]{2016ApJ...827L..30H}. Indeed, the GOES time derivative is now often used as a proxy for the HXR light curve if that is not available for a particular event, e.g.\ \cite{2008SpWea...6.5001C}, \cite{2014ApJ...787...32M}.  With the upgrades on GOES 13, 14, and 15, improved measurements of the SXR flux with 2 s time resolution and finer digitization have been available since 2010. \cite{2012ApJ...749L..16D} and \cite{2015SoPh..290.3625S} have used these new observations to show much more extensive SXR pulsations than had previously been suspected, persisting well into the decay phase in some cases and evident at other wavelengths. These pulses are best seen either in the time derivative of the SXR light curve or after subtracting the more gradually varying component.  The times between peaks generally range from the lower detectable limit of a few seconds up to $\sim$100~s.

Much effort has been expended in searching for periodicities in these pulses, and the name quasi-periodic pulsations (QPP) has been used \citep{2009SSRv..149..119N, 2016ApJ...822....7K}.  However, no convincing evidence for sustained periodicities has been found with significant power at any particular frequency.  This is the case especially when the underlying noise distribution is taken to be a power-law in frequency, so-called ``red-noise,'' as opposed to the usual assumption of ``white-noise" that is independent of frequency \citep{2011A&A...533A..61G, 2013ApJ...768...87H, 2015ApJ...798..108I, 2016ApJ...........I}.

Despite the many decades since the discovery that flares have an impulsive phase, the origin or cause of the multiple peaks in the emissions is not known.  \cite{2009SSRv..149..119N} have separated the possible mechanisms that could give rise to these pulsations into so-called ``load/unload'' mechanisms and MHD oscillations. Thus, there are two broad possibilities for the origin of the pulsations. One is that the pulsations are the result of multiple bursts of energy believed to result from repeated episodes of magnetic reconnection in a coronal current sheet or sheets, such as the multi-island reconnecting system proposed by \cite{2006Natur.443..553D}. In this case, the pulsations could reveal fundamental properties of the energy release process itself.  The second possibility is that the pulsations arise from MHD waves propagating in magnetic loops \citep{1983SoPh...88..179E, 2005LRSP....2....3N, 2016GMS...216..395W}.  Their characteristics would depend on the type of MHD waves (e.g.\ kink modes, fast sausage modes \citep{2016SoPh..291..877G, 2016ApJ...823L..16T}, or slow modes) and on the properties of magnetic loops in which they propagate, their length, density, magnetic field strength, etc. A~combination of these two scenarios is also possible but it might be expected that the time scales and relevant length scales would be quite different in the two cases -- multiple bursts of energy released in a current sheet are expected to be on shorter (sub-second {to second}) time scales and (sub-arcsecond) length scales than the characteristic scales of MHD oscillations in a magnetic loop (e.g. \cite{2003astro.ph..9505A}).

Depending on the mechanism by which the measured SXR pulsations are produced, the timescales involved most probably depend on the phase speed of the wave propagation divided by the length scale of the source. We assume initially that the relevant speed is the Alfv\'en speed, which is proportional to the magnetic field strength and inversely proportional to the square-root of the density. Thus, in principle, we can obtain estimates of all the relevant parameters in the system except for the magnetic field strength.  The length scale of the source is obtained from images made with the Ramaty High Energy Solar Spectroscopic Imager \citep[RHESSI; ][]{2002SoPh..210....3L} and the Atmospheric Imaging Assembly \citep[AIA; ][]{2012SoPh..275...17L} on the Solar Dynamics Observatory  \citep[SDO; ][]{2012SoPh..275....3P}.  The density of the SXR-emitting plasma is calculated from the emission measure derived from the 2-channel GOES flux measurements and the source volume from the RHESSI images.  Thus, we can obtain an estimate of the coronal magnetic field in the vicinity of the flare-heated plasma. This method has been used previously by \cite{2001A&A...372L..53N} and \cite{2011ApJ...736..102A} for coronal loop oscillations, which they assumed were the result of standing kink waves with a phase speed equal to $\sqrt{2}$ times the Alfv\'en speed when the plasma density outside the loop is negligible. The method is also similar to the technique used by \cite{2016NatPh..12..179J} to provide magnetic field mapping capability in the corona from measurements of sunspot oscillations. We have used this general approach to explore the consistency of the various model predictions with the observations of SXR pulsations in one well-observed event.  Clearly, more sophisticated models should be explored once the viability of the method is established and the ability of the observations to distinguish between different production mechanisms can be demonstrated.  For this paper, we assume that the Alfv\'en speed and the loop length are the relevant parameters that control the measured QPP period and its variation with time.  The implications of assuming different speeds and length scales are explored in the discussion section.

We examine the pulsations in the X3.2 flare on 2013 May 14 (SOL2013-05-14T01:11), paying particular attention to their continuation into the decay phase.  X-ray observations of this event made with RHESSI have already been analyzed by \cite{2015arXiv150702009C}. They carried out a standard Fourier and Lomb-Scargle periodogram analysis of the RHESSI light curves at different energies and found periods of $\sim$53 and $\sim$72~s.  
However, they did not consider the ``red noise'' nature of the flare Fourier spectra reported by \cite{2015ApJ...798..108I} in assessing the significance of the peaks found in the power spectra nor have they considered the variation of the periods with time.  They concluded that the QPP originate from compressible fast magnetoacoustic sausage-mode oscillations in the flaring loop system and that the two periods are the fundamental and first harmonic.

\cite{2015A&A...574A..53K} have also studied QPP from this same flare.  They used 17 GHz microwave observations from the Nobeyama Radioheliograph for just 400 s covering the impulsive phase of the flare.  Their Hilbert-Huang transform analysis revealed three intrinsic modes of oscillations with mean periods of 15, 45, and 100~s that the authors attributed to both kink and sausage modes of the flaring loop.  They also report rapidly decaying horizontal kink oscillations of neighboring loops seen in SDO/AIA movies with periods of several minutes.

For our analysis of this event, we have concentrated on the {\em GOES} SXR observations since they are capable of revealing the smallest pulsations in flux in the $\sim$1--10~keV energy range.  The solar origin of these pulsations, as opposed to any instrumental effects, can be clearly demonstrated for those events that have been observed simultaneously with instruments on both GOES 13 and GOES 15.  One such event that started on 11 March 2015 at 16:11 UT (SOL2015-03-11T16:11) shows distinct pulsations in the light curves from both instruments extending for up to 30 minutes after the peak.\footnote{
  RHESSI Nugget \#262, \emph{Fine Structure in Flare Soft X-ray Light Curves}, Dennis, B.~R.~and~Tolbert, A.~K.}
In our case, we compared the GOES light curves with the light curves from the SXR channel (0.1--7~nm or 1--70~\AA) of the EUV SpectroPhotometer \citep[ESP; ][]{2012SoPh..275..179D}, which is part of the EUV Variability Experiment \citep[EVE; ][]{2012SoPh..275..115W} on SDO. The light curves from these two instruments that cover almost identical wavelength ranges are very similar,  clearly demonstrating that the pulsations are 
of solar origin and that their properties can be studied in detail down to the 2~s time resolution of the 
GOES-15 observations.

{One complication in using the GOES or ESP data is that both instruments record the total solar flux.  Thus, X-ray pulses of interest from the X-class flare itself cannot be separated from X-ray flares in other active regions, especially those occurring during the long decay phase.  These separate flares can give peaks in the wavelet power that confuse the analysis of the QPP from the primary flare of interest.  Fortunately, they can be identified in the full-Sun images available from SDO/AIA, GOES/SXI, and Hinode/XRT, and their effects minimized in the subsequent analysis (see below).}

{In searching for QPP in the GOES and ESP light curves, we} have used the basic wavelet analysis method of \cite{1998BAMS...79...61T} with modifications described in Appendix \ref{wavelet}. We chose this method rather than the more common Fourier power-spectrum analysis since we believe that it is more appropriate for the apparent ``quasi-periodic'' nature of the observed pulsations in the GOES light curves. Fourier power-spectrum analysis is used to search for sustained periodicities in which both the period and phase are maintained during the analyzed time interval. Given the observed evolution of many flares with the progression of the apparent energy release site along an arcade of magnetic loops and to higher altitudes, it would be surprising if any single period or multiple periods were sustained throughout the duration of the flare.  Furthermore, if multiple locations or loops did have the same or similar period, it would be difficult to understand how the phase of any oscillation could be maintained from one location to the next during a flare.  These concerns are in addition to the form of the power spectrum itself with the background power being either independent of frequency (``white noise'') or following a power-law dependency on frequency (``red noise'') as mentioned above.

Wavelet analysis, in contrast to Fourier analysis, allows power to be evaluated over a broad range of time scales as a function of time throughout the flare.  No periodicity is required to generate wavelet power, just variations in the flux on the time scales covered by the wavelet durations.
This technique has allowed us to detect significant pulsations extending well into the decay phase of the event, and to identify trends in the timescales of the pulsations as a function of time.  We have related these results on the pulsation timescales to source intensities, dimensions, and locations determined from RHESSI and AIA images with a view to determining the mechanism by which these pulsations are produced.

It has long been known that the decay time of SXR events is much longer than the expected conductive and cooling times of the hot plasma heated during the impulsive phase.  While this could be explained in some cases by the suppression of thermal conduction \citep{2015ApJ...811L..13W}, it is generally attributed to additional heating that continues after the impulsive phase.  This is presumably the result of magnetic reconnection at successively higher altitudes that produces the bright post-flare loops seen in H-alpha and in various UV and EUV wavebands.  The continuation of SXR pulsations into the decay phase could also be a manifestation of this continued heating, and they could be useful in determining the characteristics of the processes involved.

The flare observations and data analysis are described in Section 2. The wavelet analysis is described in Section 2.2 and Appendix A.  The analysis of RHESSI and AIA images is presented in Section 2.3.  Section 3 contains an interpretation of the results leading to an evaluation of the coronal magnetic field strength based on an MHD model.  Section 4 has a summary of the results and an evaluation of future requirements.



\section{OBSERVATIONS AND DATA ANALYSIS}
\subsection{Flare Selection and Time Coverage}
\label{flareselection}

\begin{figure*}
    \includegraphics*[angle = 90,width = 1.0\textwidth]
    {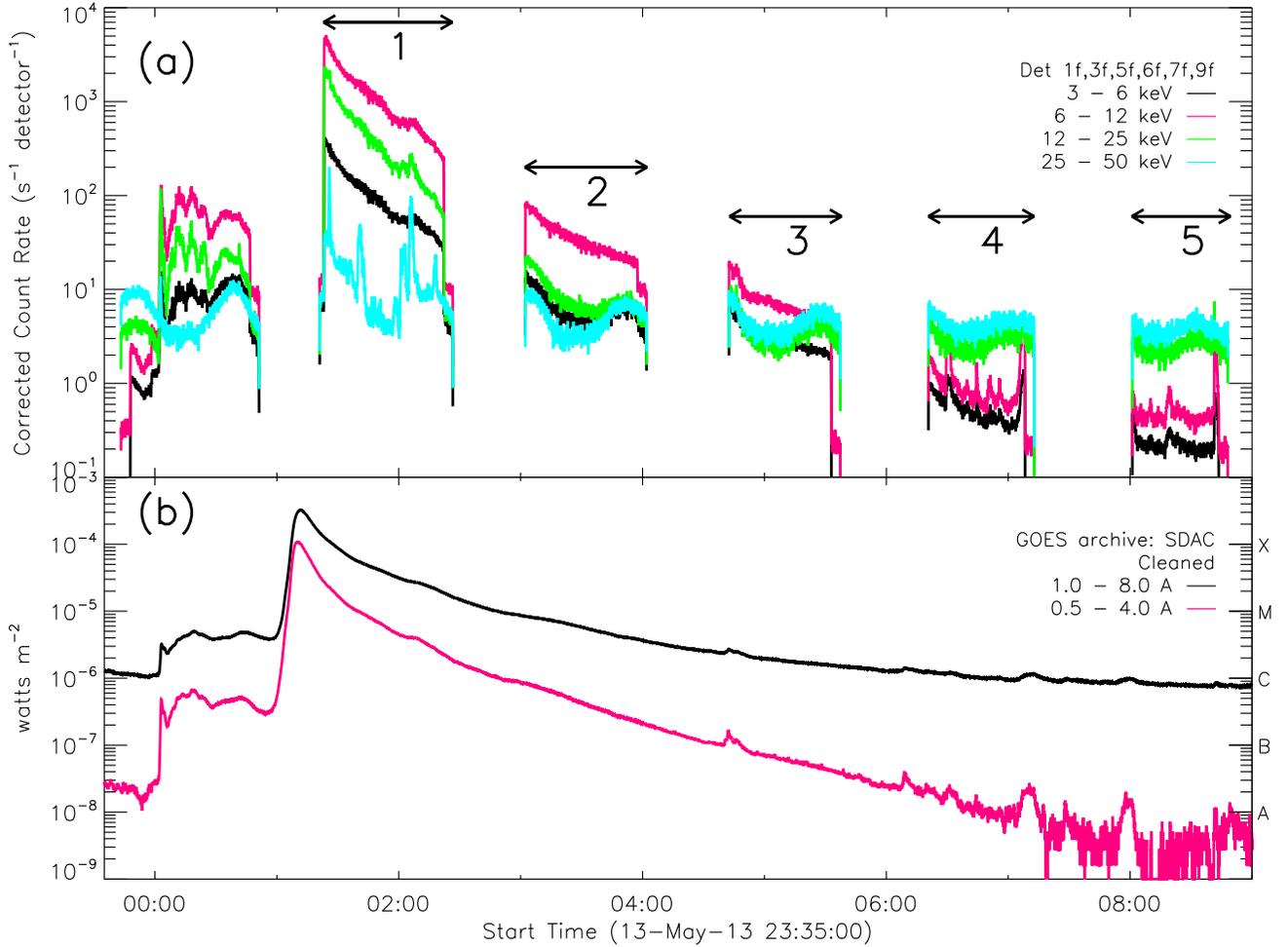}
    \caption{
    {\em RHESSI} and {\em GOES} light curves for the X3.2 flare on 2013 May 14.
    (a) RHESSI count-rate light curves corrected for the different attenuator states (A1 in orbits 1 and 2, A0 in orbits 3, 4, and 5) and plotted in 4~s time bins and four energy bands: $3-6$~keV (black), $6-12$~keV (pink), $12-25$~keV (green), $25-50$~keV (cyan). The gaps result from the $\sim$35~minute nighttime part of RHESSI's 96-minute orbit.  The five orbits of observations used in this paper are labeled starting with the first orbit after the soft X-ray peak at 01:11~UT, which occurred during RHESSI night.  The higher energies returned to background levels in Orbit 2 but the 3--6 and 6--12 keV rates were above background, and images could still be made, as late as Orbit 5.
    (b) {\em GOES--15} X-ray flux light curves at 1--8~\AA~(black) and 0.5--4~\AA~(pink) with 2~s time cadence.
    }
    \label{fig:lc_hsi_goes}
\end{figure*}

We chose for analysis the X3.2 event on 2013 May 14 that occurred in NOAA active region 11748 near the north-east limb at N13E80.  The flare started in the GOES 1--8~\AA~light curve at 00:00~UT, peaked at 01:11 UT, and returned to the preflare background level at about 08:00 UT (Fig.~\ref{fig:lc_hsi_goes}).  
RHESSI was at night during the impulsive phase but recorded the preflare emission for one orbit starting at 00:00 UT and the full decay phase after 01:20 UT with regular nighttime interrupts for an additional five orbits. The Fermi Gamma-ray Burst Monitor \citep[GBM;][]{2009ApJ...702..791M} observed the full impulsive phase with observations starting at 01:02 UT until nighttime at 02:02 UT. However, severe count-rate saturation limits the use of the GBM observations, and they have not been considered further for this analysis.

\begin{figure}
    \includegraphics*[angle = 0,width = 0.5\textwidth]
    {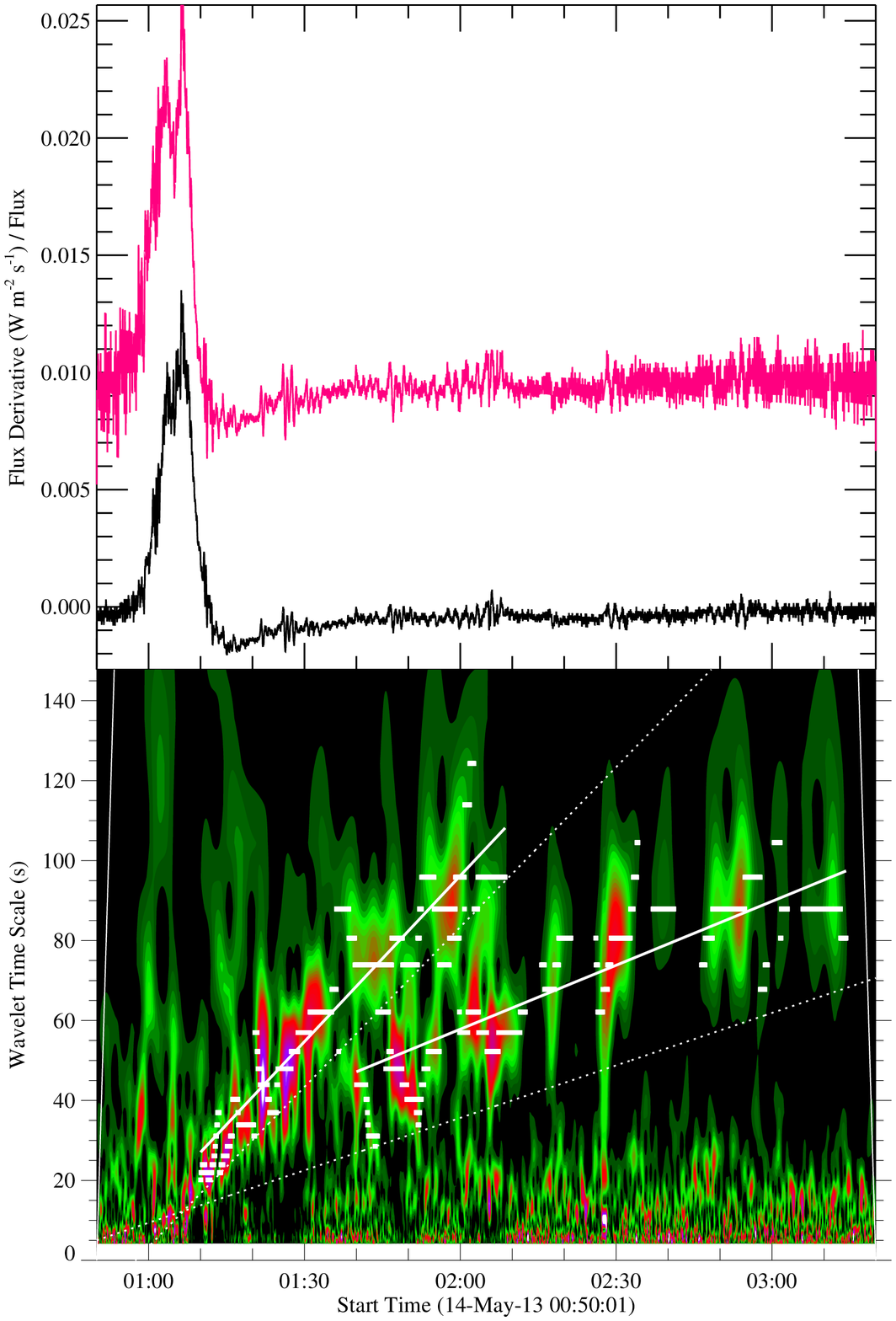}
    \caption{{\em GOES} fractional time derivatives and wavelet power. Top panel: Time derivatives plotted with 2~s cadence of the {\em GOES}~$1-8$~\AA~(pink, offset by 1\%) and $0.5-4$~\AA~(black) fluxes as a fraction of the flux in that channel. Bottom panel: Normalized wavelet power per second of the {\em GOES}~$1-8$~\AA~light curve plotted as a function of wavelet time scale and time. The two diagonal solid white lines show linear fits to the positions of peak power at each time interval (shown as short horizontal white bars) the ranges between and above the time scales defined by the two dotted lines. The peaks below the lower dotted line are from the statistical and digitization noise in the data.
      }
    \label{fig:goes_lc_deriv_wavelet}
\end{figure}

The reason for selecting this event for analysis is that the pulsations seen in the time derivative of the GOES light curves continue for over two hours after the impulsive phase.  This is shown in Fig.~\ref{fig:goes_lc_deriv_wavelet}, where the fractional time derivatives of the {\em GOES} 1--8~\AA~and 0.5--4~\AA~fluxes - $(dI_i/dt)/I_i$ - are plotted vs.\ time (note the 1\% offset of the 1--8~\AA~channel in the vertical direction to separate the two channels). Here, $I_i$ is the flux ($W~m^{-2}$) and $dI_i/dt$ is the time derivative in GOES channel $i$. The initial peak in this plot at 01:06~UT shows the impulsive rise of the flux at a rate of $\sim$1.6\% of the flux per second. The derivative decreases to zero at 01:11 UT, the time of the SXR maximum flux, and then becomes negative as the flux decreases. The structures in the plot {(apart from some of the pulses after 01:45 UT that are discussed below and appear to be from small flares in a different active region )} are the quasi-periodic pulsations that are the subject of this paper.  The noise level is negligible except near the end when the digitization step size becomes comparable to the amplitude of the fluctuations (see \cite{2015SoPh..290.3625S} for a discussion of the noise in the GOES data).

\begin{figure}
    \includegraphics*[angle = 0,width = 0.5\textwidth]
    {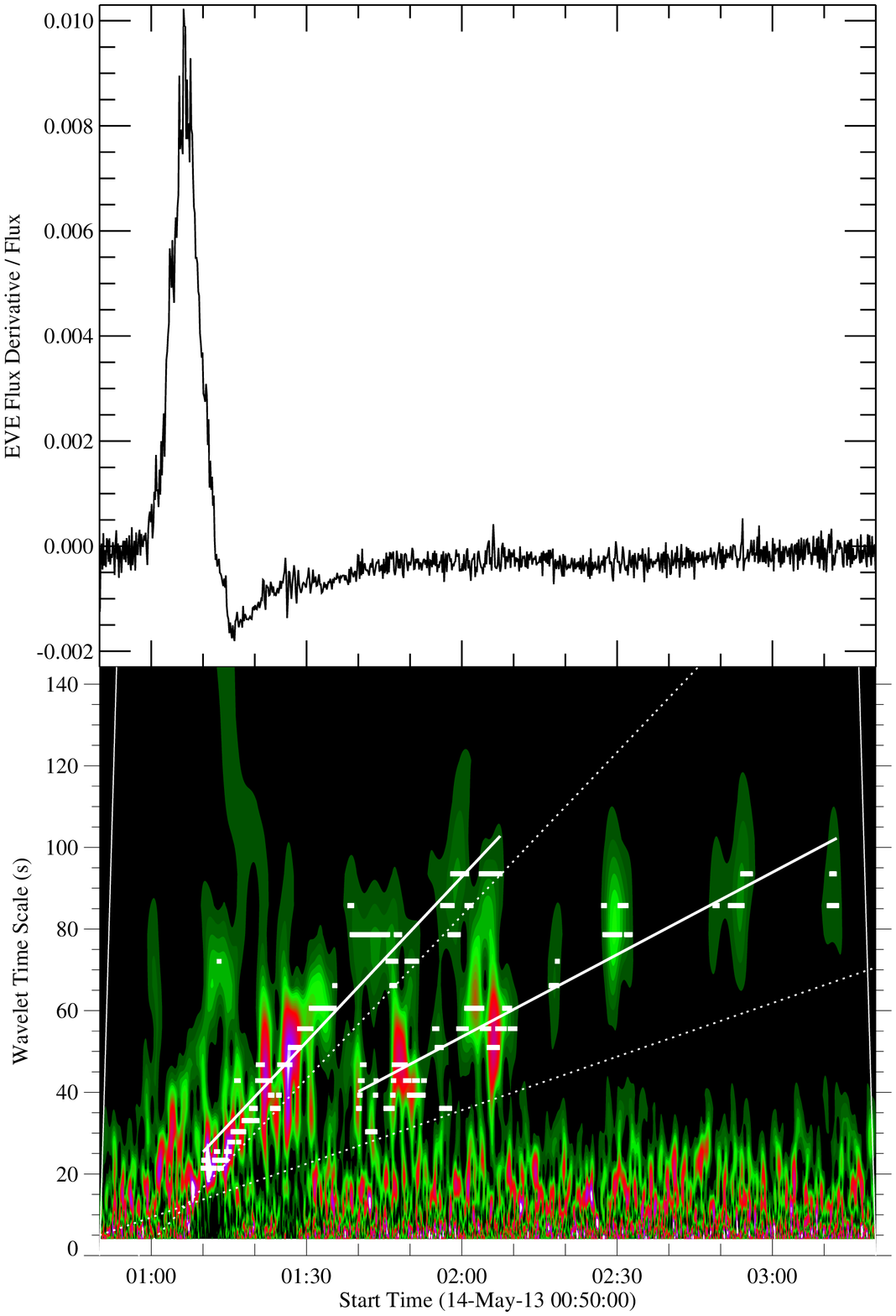}
    \caption{Same as Fig.\ \ref{fig:goes_lc_deriv_wavelet} for the light curve of the EVE/ESP 1--70~\AA~flux.}
    \label{fig:esp_lc_deriv_wavelet}
\end{figure}

The time derivative of the EVE/ESP 1--70\AA~flux is shown in Fig.\ \ref{fig:esp_lc_deriv_wavelet} for direct comparison with the GOES time derivative in Fig.\ \ref{fig:goes_lc_deriv_wavelet} covering the same time period.  The two light curves are remarkably similar and both show the same, almost identical, pulsations.  The similarity between the two light curves is borne out by the detailed wavelet analysis of the two data sets as discussed below. The fact that the same fluctuations can be seen in observations made by two instruments covering similar wavelength ranges but on different spacecraft shows that the pulsations are most likely of solar origin and not the result of some instrumental or data analysis artifact.

\subsection{Wavelet Analysis}

In order to identify and characterize the pulsations seen in the time derivative plots of Fig.\ \ref{fig:goes_lc_deriv_wavelet}, we have used the IDL wavelet software discussed by \cite{1998BAMS...79...61T} and modified as described in Appendix \ref{wavelet}.  The emphasis in the analysis was to optimize the procedure for the detection of pulsations with periods between a few seconds and a few hundred seconds. In the process, information about the amplitudes of the the pulsations was sacrificed. The final product shown at the bottom of Fig.\ \ref{fig:goes_lc_deriv_wavelet} is the normalized wavelet transform power plotted vs. time and wavelet timescale.  This plot covers the interval from the start of the impulsive phase of the main event at 00:50~UT until 03:20~UT, when the wavelet power became undetectable above the readout noise.  It reveals that the time scale of the fluctuations in the time derivative light curves changed systematically with time.  The two solid white lines show linear fits to the variation of time scales with peak power in two ranges delineated by the white dotted lines.  No significant wavelet power is evident after 03:20~UT except for the contribution from a small peak in the GOES light curve (Fig.\ \ref{fig:lc_hsi_goes}(b)) between 04:40 and 04:50 UT, which is from a separate flare as discussed in Section \ref{Imaging}.

\begin{figure}
    \includegraphics*[angle = 0,width = 0.5\textwidth]
    {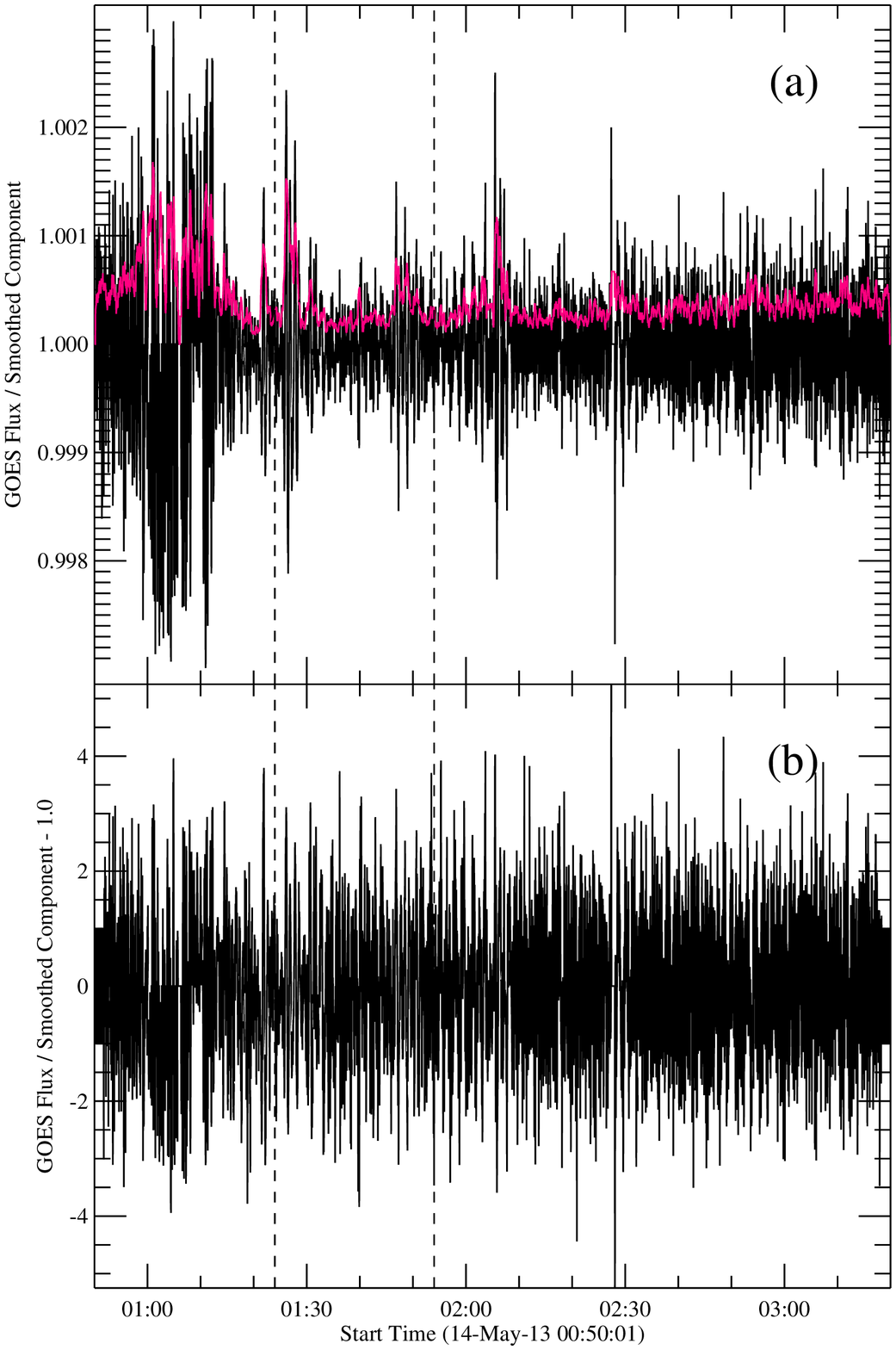}
    \caption{High frequency components of the {\em GOES}~$1-8$~\AA~light curve. (a)~The black curves show the GOES flux divided by the flux smoothed with a box car that is 10 2-s time intervals wide.  The red curve is the absolute value of the black curve smoothed with a box car that is 10 2-s time intervals wide. (b)~The normalized relative intensity obtained by dividing the black curves in the top plot by the red curve and subtracting 1.0. The two vertical dashed lines show the time interval plotted on an expanded scale in Fig.~\ref{fig:lc_hsi_goes_desmoothed}.}
    \label{fig:lc_goes_desmoothed}
\end{figure}

\begin{figure}
    \includegraphics*[angle = 0,width = 0.5\textwidth]
    {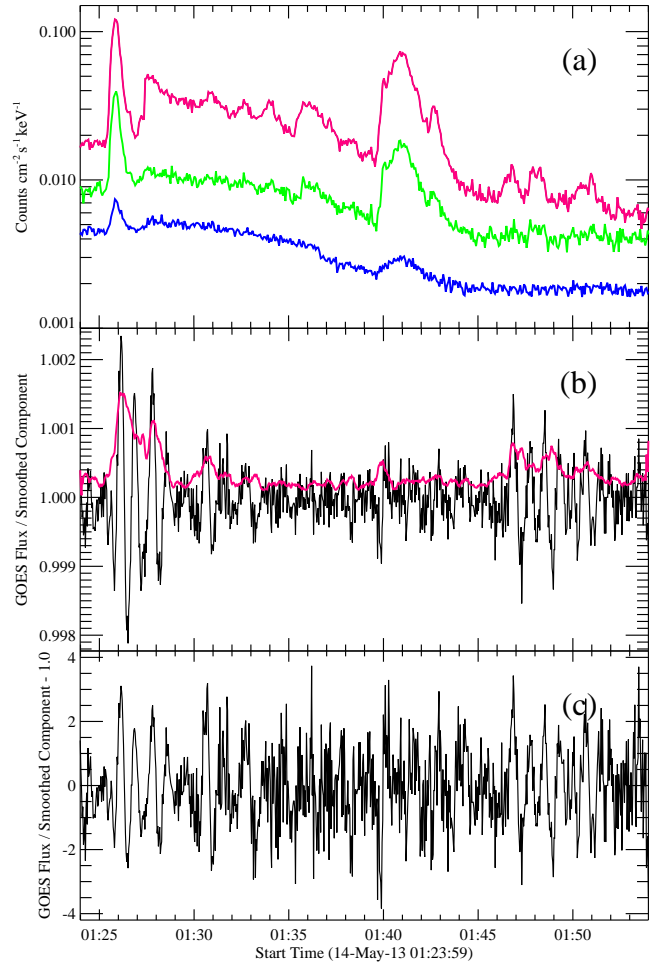}
    \caption{Expanded light curves for the 30 minute time interval between the vertical dashed lines shown in Fig.\ \ref{fig:lc_goes_desmoothed}. (a) RHESSI count flux light curves in the 25--50~keV (red), 50--100~keV (green), and 100--300~keV (blue) energy bands plotted with 4 s cadence. (b) and (c)  Same {\em GOES}~$1-8$~\AA~light curves as in Fig.\ \ref{fig:lc_goes_desmoothed} showing the high frequency components.}
    \label{fig:lc_hsi_goes_desmoothed}
\end{figure}

The starting point for generating the wavelet transform power plot shown in Fig.\ \ref{fig:goes_lc_deriv_wavelet} was the {\em GOES--15}~1--8~\AA~light curve shown in Fig.\ \ref{fig:lc_hsi_goes}(b). The steps taken to obtain this plot and the rationale for each step are as follows:

\begin{enumerate}

\item The GOES fluxes measured every 2 s were divided by values obtained from a box-car smoothed version of the light curve with a box-car width of 20~s.  This step was necessary to remove the gradually varying component so that the final wavelet transform power would better show the more rapidly varying component that was of greater interest. The result is shown as the black curve in Fig. \ref{fig:lc_goes_desmoothed}(a) and on an expanded time scale in Fig.\ \ref{fig:lc_hsi_goes_desmoothed}(b). It has both positive and negative excursions from the average value of~1.0.
\item The absolute value of the result from Step 1 was again smoothed using a box car with a width of 20~s. The result is shown as the red curves in Figs.~\ref{fig:lc_goes_desmoothed}(a) and \ref{fig:lc_hsi_goes_desmoothed}(b).
\item The result from Step 1 was divided by the result from Step 2 and, after subtracting 1.0, the result is shown in Figs.~\ref{fig:lc_goes_desmoothed}(b) and \ref{fig:lc_hsi_goes_desmoothed}(c). This step was used so that the wavelet transform power would be more dependent on the distribution of time scales at any given time rather than on the amplitude of the signal.  This allows significant pulses to be detected over a wide range of amplitudes from the largest during the impulsive phase to the smallest in the decay phase.
\item The result from Step 3 was input to the IDL wavelet software discussed by \cite{1998BAMS...79...61T}.  The Morlet wavelet was selected and the complex wavelet transform computed as a function of the wavelet time scale and time.
\item 
    The wavelet transform power per second was computed as the square of the absolute value of the wavelet transform divided by the wavelet time scale at which the wavelet transform is calculated, i.e., $|W_{i,j}|^{2}~/~\sigma_{j}$ where $W_{i,j}$ is the wavelet transform at time $i$ and wavelet scale $j$; $\sigma_{j}$ is the wavelet time scale.
    The division by the wavelet time scale was required to account for the different durations of the wavelet as a function of timescale - see Appendix \ref{wavelet}.
\end{enumerate}

The normalized wavelet transform power per second obtained in Step 5 is plotted in Fig.\ \ref{fig:goes_lc_deriv_wavelet}. It shows two ``ridges'' where the position of peak power moves in time to longer timescales.  The first ridge from the X-class flare starts near the peak of the event at 01:10~UT with a timescale of $\sim$25~s and trends upwards to reach $\sim$85~s at 01:55~UT with a rate of change \boldmath{$\dot{\tau} = 2.2 \times 10^{-2}$}.  {The second ridge is likely to have resulted from small flares in a different active region as discussed below. It }starts at $\sim$01:40~UT with a period of $\sim$45~s and ends at $\sim$03:10~UT with a period of $\sim$90~s and \boldmath{$\dot{\tau} = 8.3 \times 10^{-3}$}. The two white lines in Fig.~\ref{fig:goes_lc_deriv_wavelet} are linear least-squares fits to the peaks of these two ridges.

An identical wavelet analysis was carried out on the EVE/ESP observations of this event.  The resulting normalized wavelet transform power per second is shown in Fig.\ \ref{fig:esp_lc_deriv_wavelet}.  Comparison with the equivalent plot for the GOES data in Fig.\ \ref{fig:goes_lc_deriv_wavelet} shows the same structure with the same two ridges of high power demonstrating that these features must be of solar origin and not some instrumental artifact.

\begin{figure}
    \includegraphics*[angle = 90, width = 0.5\textwidth]
    {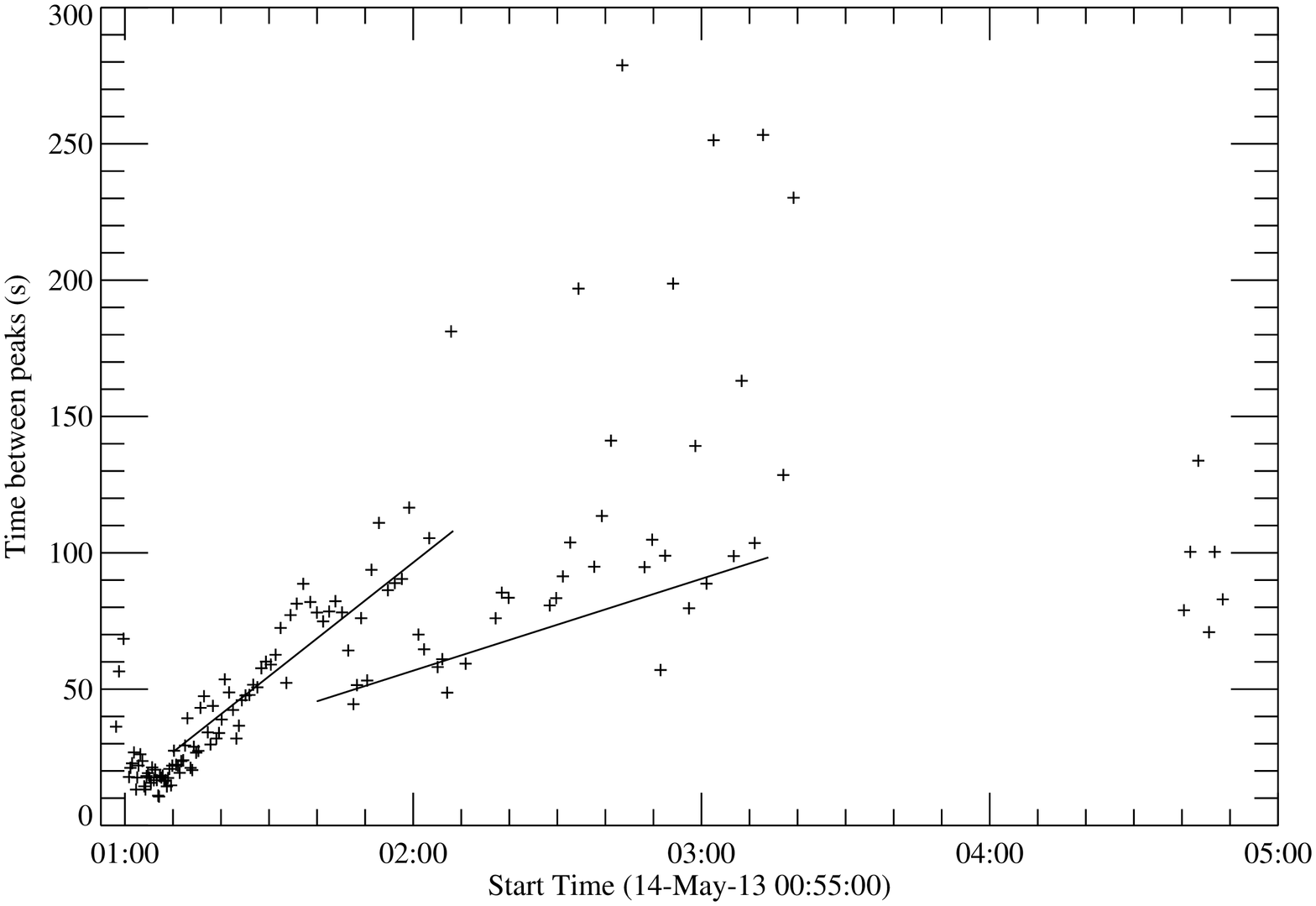}
    \caption{Times between neighboring peaks identified by eye in the GOES time derivative light curve shown in the top plot of Fig.\ \ref{fig:goes_lc_deriv_wavelet}.  The two solid lines correspond to the linear fits to the peak wavelet transform power shown in the lower plot of Fig.\ \ref{fig:goes_lc_deriv_wavelet}. Times between peaks of $>$300~s are not shown. Note that the 6 points after 04:40 UT are for a small sub-flare during the long decay of the main event (see Section \ref{Imaging}).}
    \label{fig:time_between_GOES_peaks}
\end{figure}

We have checked the validity of the timescale trends found from the wavelet analysis by comparing them to the peaks visible in the GOES time derivative plot shown in Fig.\ \ref{fig:goes_lc_deriv_wavelet} and especially in the plots of Figs.\ \ref{fig:lc_hsi_goes_desmoothed}(b) and (c) with the expanded timescale.  We determined the time of all 163 peaks identified by eye in the time-derivative light curve and then plotted the times between adjacent peaks vs.\ time shown in Fig.\ \ref{fig:time_between_GOES_peaks}.  The time between most peaks is similar to the timescale with peak wavelet power at any given time.  Furthermore, the points cluster along two trend lines similar to the ones shown by the two solid lines taken from Fig.\ \ref{fig:goes_lc_deriv_wavelet}.  The close agreement between the wavelet power trend lines and the times between peaks shows that the wavelet analysis is picking up the local timescale of the signal as expected. As with the wavelet analysis, no significant peaks were identified by eye after 03:20~UT except for the peaks between 04:35 and 04:50~UT that were from a separate small flare {from the same active region} as discussed in Section \ref{Imaging}. This comparison serves to validate the wavelet analysis for this event and suggests that it can be used for the analysis of other events where the individual peaks may not be so clearly apparent above the noise in the data.

The RHESSI light curves in Fig.\ \ref{fig:lc_hsi_goes_desmoothed}(a) show significant peaks at energies between 25 and 300 keV but they are not consistently related to the pulsations seen in the GOES light curves shown below in Figs.\ \ref{fig:lc_hsi_goes_desmoothed}(b) and (c).  The prominent HXR peak at 01:26~UT, for example, is accompanied by at least three large pulsations in the GOES light curve but the HXR peak at 01:40 UT does not have an obvious GOES counterpart.  The detailed comparison of pulsations in SXR and HXR flare light curves is the subject of a companion paper in preparation \citep{2016ApJ...827L..30H} so this issue is not pursued further here. {Also, see Section \ref{Other ARs} for a discussion of the effects of small flares from other active regions on these light curves.}

\subsection{Imaging Analysis}
\label{Imaging}

\begin{figure} \centering
    \includegraphics*[angle = 0, width = 0.5
    \textwidth, trim = 0 0 150 0]
        {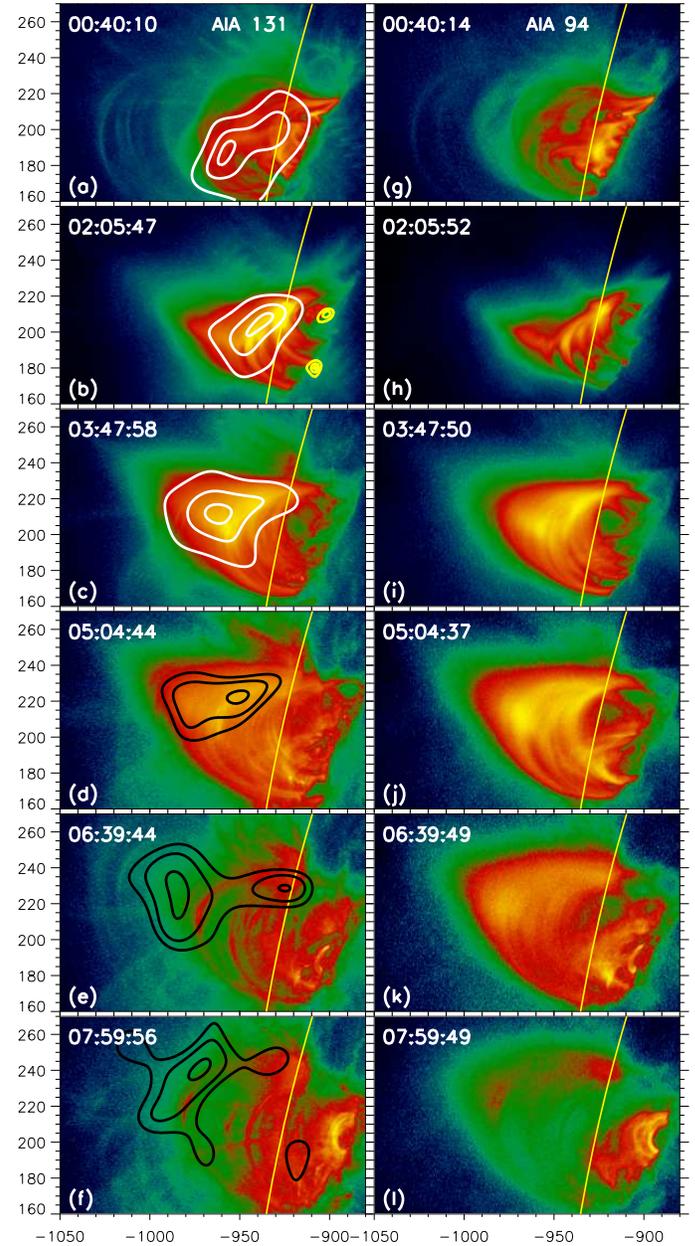}
    \caption{{\em AIA}~131\AA~(left column) and 94\AA~(right column) images taken near the mid-point of each of the six RHESSI orbits shown in Fig.~\ref{fig:lc_hsi_goes}. The following RHESSI contours are overlaid on the 131 \AA~ images: black - 3--6~keV, white - 6--12~keV, and yellow - 25--50~keV. The contour levels are at 10, 50, and 90\% of the peak intensity in each image.}
    \label{fig:im6_AIA_RHESSI}
\end{figure}

The near-limb location of the flare (N13E80) allows the altitude of the SXR source and the loop lengths to be readily determined from RHESSI and AIA images.  To provide an overview of the flare evolution, RHESSI images were made in four energy ranges (3 - 6, 6 - 12, 12 - 25, 25 - 50 keV) for the six orbits identified in Fig.\ \ref{fig:lc_hsi_goes}, when the emission was detectable above background between 00:00 UT and 08:40~UT. The images were made with the visibility-based maximum entropy method called MEM NJIT \citep{2007SoPh..240..241S} using detectors 4 through 9. The resulting contours are shown in Fig.\ \ref{fig:im6_AIA_RHESSI} for six times near the mid-points of the six RHESSI orbits overlaid on the AIA 131~\AA~images taken near the same times.

\begin{figure*} \centering
    \includegraphics*[angle = 90, width = 1.0\textwidth]
    {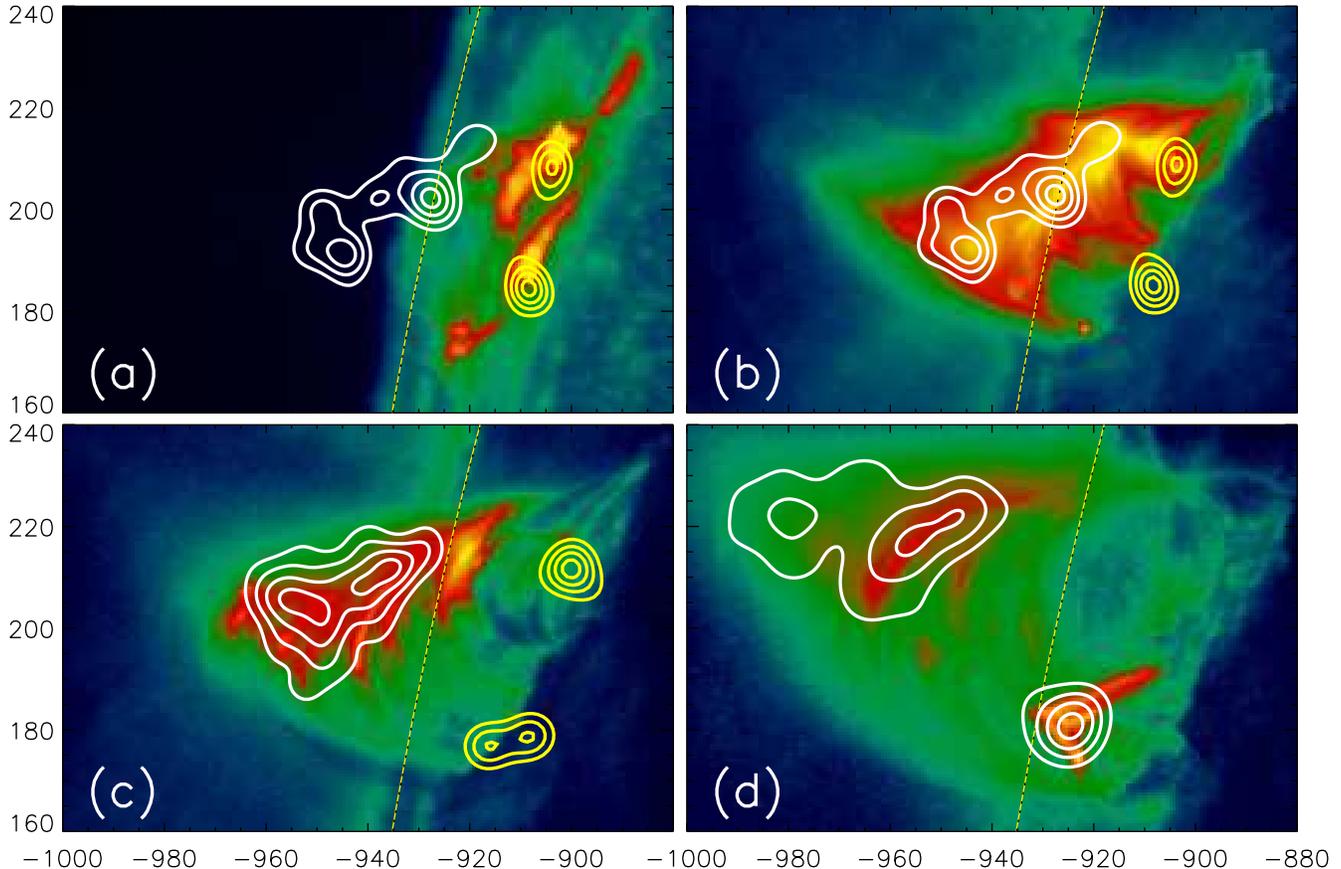}
    \caption{{\em AIA}~images at four times during the flare overlaid with RHESSI contours of fluxes in the 6 - 12 keV (white) and 25 - 50 keV (yellow) energy ranges.  The contours are at 30, 50, 70, and 90\% of the peak value in each energy range. (a) {\em AIA}~1700~\AA~image at 01:26:07 UT with RHESSI contours for the period from 01:25:32 to 01:26:16 UT, during a peak in the 25--50 keV light curve near the start of the first orbit identified in Fig.\ \ref{fig:lc_hsi_goes}. (b) {\em AIA}~193~\AA~image at 01:26:09 UT with the same RHESSI contours as in (a). (c) {\em AIA}~193~\AA~image at 02:18:09 UT with RHESSI contours for the period from 02:17:08 to 02:18:56 UT, near the end of the first RHESSI orbit. (d) {\em AIA}~131~\AA~image at 04:47:44 UT with RHESSI 6--12~keV contours for the period from 04:47:20 to 04:51:56~UT near the start of the third orbit. Note that a C-class flare produced the bright cusped loops on the solar disk in this AIA image and the X-ray source from the same location.  The thermal coronal emission from the long-lasting X3.2 flare is still visible above the limb in both the AIA and RHESSI images.}
    \label{fig:im4_AIA_RHESSI}
\end{figure*}

Fig.~\ref{fig:im4_AIA_RHESSI} shows four {\em AIA} images at different times during the flare with RHESSI contours overlaid. In this case, the RHESSI images were made with the CLEAN reconstruction algorithm \citep{2002SoPh..210...61H} using detectors 3 through 9. Fig.~\ref{fig:im4_AIA_RHESSI}(a) shows the {\em AIA}~1700~\AA~image at 01:26:07 UT, when RHESSI first started to observe this event.  The two RHESSI 25--50 keV compact sources (yellow contours) are clearly seen at footpoints since they match closely the locations of the two bright (saturated) sources in the {\em AIA}~1700~\AA~image and lie along the two ribbons seen in this image. (The $\sim$4~arcsec.\ offset in the north--south direction is probably the result of uncertainty in the AIA aspect information.) The RHESSI 6--12 keV contours (white) show a double source extending above the limb. Fig.~\ref{fig:im4_AIA_RHESSI}(b) shows an {\em AIA}~193~\AA~image at the same time with multiple bright loops also extending above the limb, matching the two RHESSI 6--12 keV sources. Fig.~\ref{fig:im4_AIA_RHESSI}(c) shows an {\em AIA}~193~\AA~image at 02:18:09 UT, close to the time of an impulsive peak in the RHESSI 25--50 keV light curve.  It also shows many bright loops now extending to higher altitudes above the limb, matching the locations of the RHESSI 6--12 keV emission, still seen as a double source. 

\begin{figure}
    \includegraphics*[angle = 90,width = 0.55\textwidth]
    {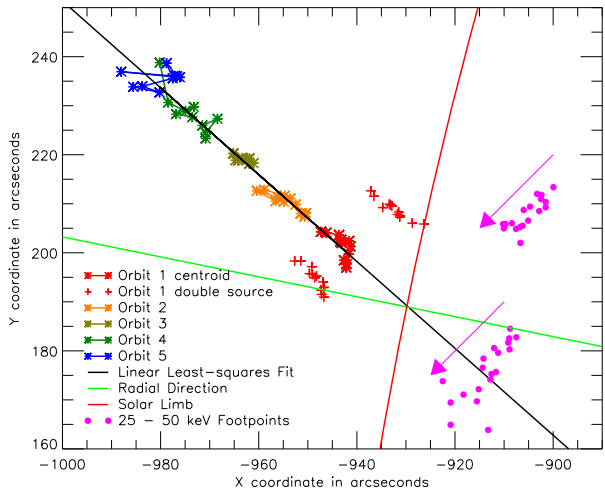}
    \caption{Location of the RHESSI source centroids from the first observation of the event at 01:24~UT during Orbit \#1 to late in the decay phase at 08:40~UT in Orbit \#5. (See Fig.\ \ref{fig:lc_hsi_goes} for orbit times.) Images in the 6--12~keV band were used up to the fourth orbit and the 3--6~keV images were used in Orbit~5. The centroid locations (asterisks) are plotted for each 5-minutes of the RHESSI observations.  The general path of the source at a position angle (measured anticlockwise from North) of $48.5^\circ$ is shown in purple as the linear least-squares fit to the centroid data points.  The green line is a radial for reference. The three sets of points for Orbit 1 reflect the double nature of the source at these times and show the locations of the centroid of all the emission taken together and of the the north and south components separately. The location of the 25--50~keV footpoints during the RHESSI Orbit 1 are shown in pink with the arrows indicating the direction of the change in position with time.}
    \label{fig:hsi_source_location}
\end{figure}

We could not verify the double nature of the source as opposed to a single extended source after the end of RHESSI Orbit \#1.  Consequently, the locations of two sources are shown in Fig.\ \ref{fig:hsi_source_location} only for Orbit \#1.

Fig.~\ref{fig:im4_AIA_RHESSI}(d) shows an {\em AIA}~131~\AA~image at 04:47:44 UT at the time of a small peak at the C2.7 level in the GOES light curves (Fig.\ \ref{fig:lc_hsi_goes}) during the decay phase of the X3.2 flare.  The coronal X-ray source of interest here appears as the extended emission from above the limb in both the AIA and RHESSI images, but note the small bright cusped loops on the solar disk in the AIA image and the source seen at the same location in the RHESSI 6--12 keV image made after coming out of night at 04:47 UT. This is a small flare that produced seven peaks in the GOES time derivative light curve between 04:35 and 04:50~UT, with the times between peaks plotted in Fig. \ref{fig:time_between_GOES_peaks}. The information about this small flare was not included in the following analysis.

Comparison with STEREO-B images from the Extreme Ultraviolet Imager (EUVI) \citep{2004SPIE.5171..111W} show that the whole active region was on the visible disk as observed from Earth orbit.  Furthermore, the RHESSI HXR footpoints were $\sim$$10^\circ$ inside the limb. Thus, it is likely that GOES, SDO, and RHESSI saw emissions from all of the event, with little if any occultation.  In particular, all the loops seen above the limb in AIA images can be confidently expected to have had their footpoints on the visible disk.  Consequently, the highest altitude of the loops at the end of RHESSI's first orbit at 02:20 UT can be computed from the $\sim60$~arcsec separation between the footpoints and the most easterly of the RHESSI 6--12 keV contours or the bright loops seen at 193~\AA\ in Fig.~\ref{fig:im4_AIA_RHESSI}(c). Assuming that the bright loops were vertical, then the maximum altitude at this time was approximately $60\times725~/~cos(10^\circ)~km = 44~Mm$.

%

The steady increase in altitude with time can be inferred from the plot of SXR source location in Fig.\ \ref{fig:hsi_source_location}.  The centroid locations plotted in this figure were obtained from the RHESSI CLEAN images. Note that the centroids of the two sources seen in the Orbit~1 images of Fig.~\ref{fig:im4_AIA_RHESSI}(a) and (b) are shown separately in addition to the centroid of all the emission above the limb considered as a single source. The locations of the 25--50~keV footpoints at various times during RHESSI's first orbit of observations from 01:26~UT to 02:20~UT are also shown.

 The average speed of the X-ray source apparent motion along the linear path shown in Fig.~\ref{fig:hsi_source_location} is $1.7~km~s^{-1}$.  This is close to the same speed determined by \cite{2002SoPh..210..341G} for a similar long-lasting X1.5 flare on 21 April 2002.

The position angle of the path of the RHESSI centroid location is $48.5^\circ$ measured counter-clockwise from solar north.  This can be compared with the position angle of the coronal mass ejections (CMEs) associated with this flare. According to the CDAW CME catalog,\footnote{http://cdaw.gsfc.nasa.gov/CME\_list/} two CMEs were observed by LASCO during the time of this flare.  The first was projected to start at 01:00 UT  with the first appearance in LASCO C2 at 01:25:51 UT. It had a linear speed of 2625~$km~s^{-1}$ and was reported to be a halo event although the brightest components were visible off the north-east limb and the mean position angle is recorded as $89^\circ$.  A second CME, first seen in LASCO C2 at 05:48:50~UT and projected to start at 05:10~UT, had a much slower speed of 672 $km~s^{-1}$ but a position angle of $62^\circ$, i.e. close to the path of the RHESSI 6--12 keV source. This is surprising since it would be expected that the reconnection of magnetic field lines dragged out by the first CME would lead to the energy release responsible for heating the plasma seen in X-rays with RHESSI.  It appears instead that the path taken by the 6--12 keV source followed more closely the direction of the northern leg of the first CME, and that this was close to the path taken by the second CME.

\subsection{{Flares from other active regions}}
\label{Other ARs}

\begin{figure}
    \includegraphics*[angle = 90, width = 0.55\textwidth]
    {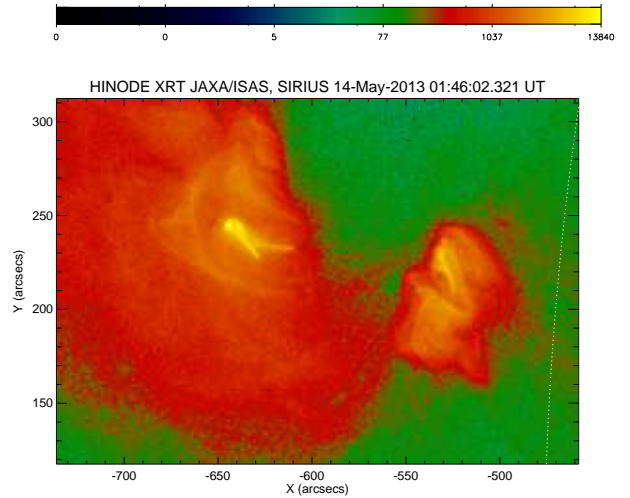}
    \caption{{Hinode/XRT image at 01:46:02 UT showing the flare in AR11745 centered at X = -640" and Y = 240."}}
    \label{fig:XRT_image}
\end{figure}

\begin{figure}
    \includegraphics[angle = 00, width = 0.5\textwidth]
	{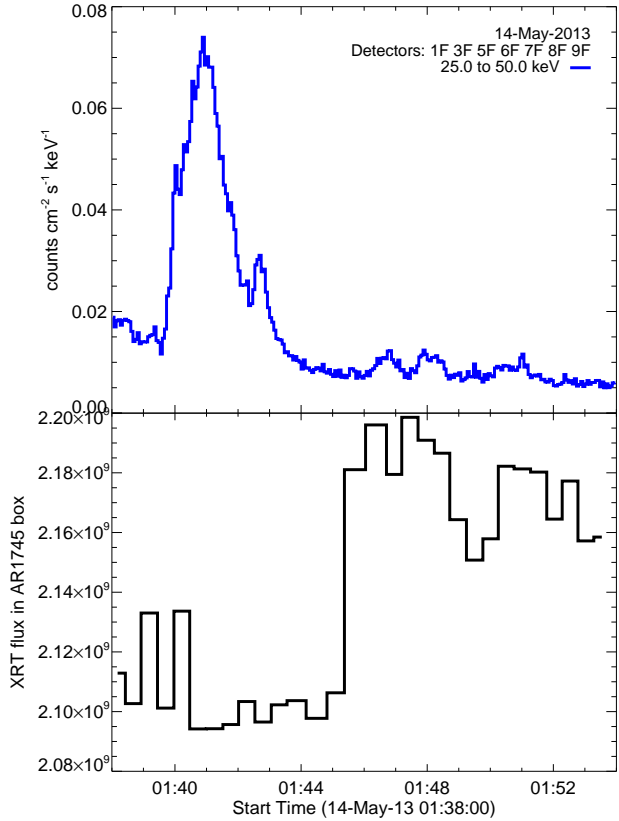}
    \caption{{Top: RHESSI 25--50 keV spatially integrated light curve. Bottom: Hinode/XRT light curve of emission from AR11745 showing a flare starting at 01:45 UT. A HXR burst from AR11748 occurs between 01:40 and 01:44 UT but there are only much weaker RHESSI signals from the burst in AR11745 seen with XRT that starts some 6 minutes later.}}
    \label{fig:XRT_lc}
\end{figure}

\begin{figure*}
    \includegraphics*[angle = 90, width = 0.8\textwidth]
    {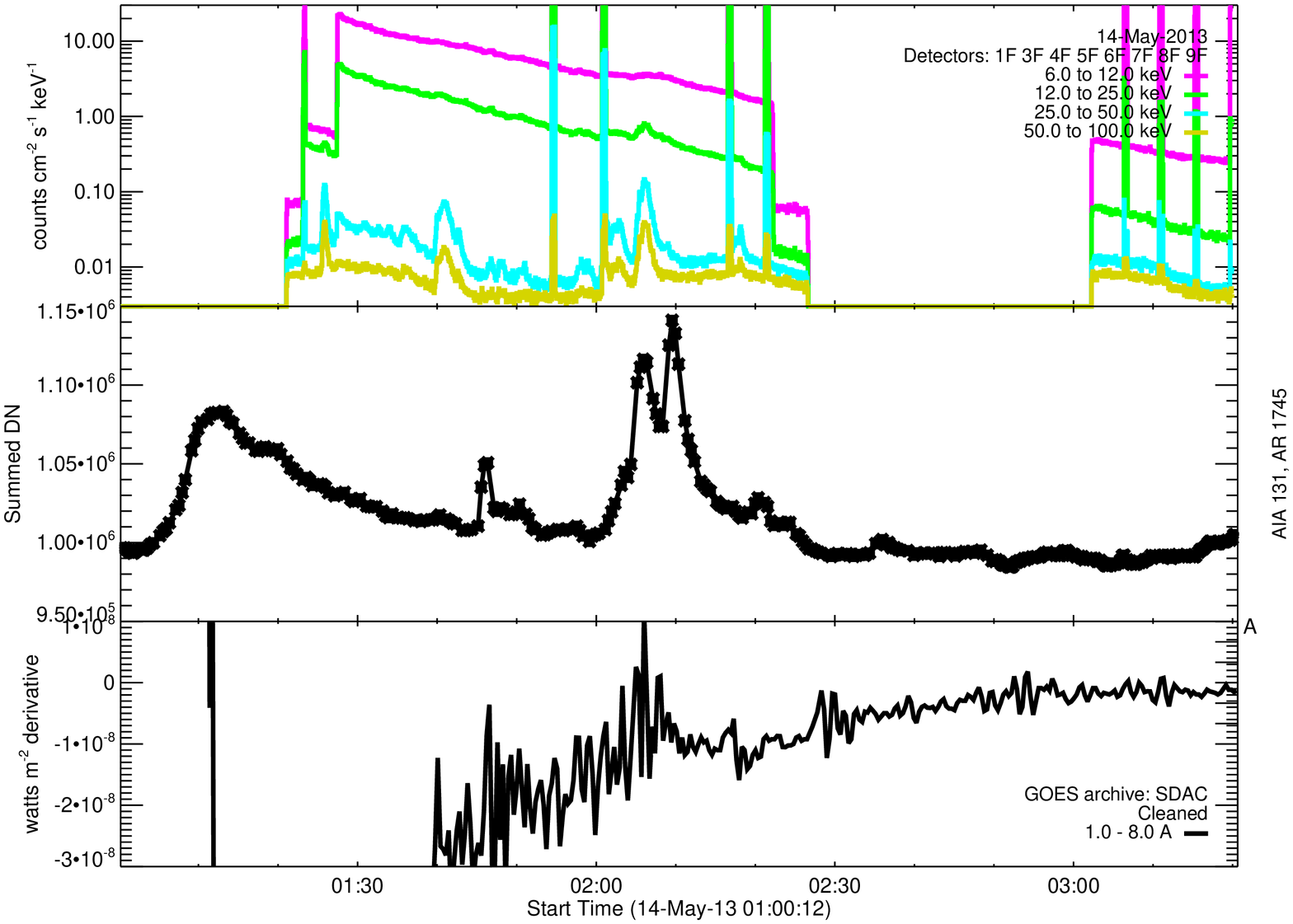}
    \caption{{Light curves from RHESSI, AIA, and GOES between 01:00 and 03:32 UT. Top: RHESSI count flux (uncorrected for attenuator changes) plotted with 4 s cadence in the 6--12, 12 - 25, 25 - 50, and 50 -- 100 keV energy ranges.  Middle: AIA 131~\AA data numbers (DNs) summed over a 200 x 200 arcsec region of interest (ROI) covering AR11745 and plotted with 12 s cadence. The gradual peak at $\sim$01:11~UT arises because of counts from the saturated source in AR11748 that leak into this ROI. The impulsive peak at 01:46~UT and the two peaks between 02:05 and 02:10~UT are from small flares in AR11745.  Note that they occur some few minutes after peaks in the HXR fluxes seen with RHESSI from AR11748 suggesting that they may be from sympathetic flares triggered by the earlier events in the neighboring active region. Bottom: GOES 1 - 8~\AA~light curve time derivative showing that the amplitude of the QPP decreases dramatically at $\sim$02:08~UT, consistent with the end time of the main ridge in the wavelet power plots of Figs.~\ref{fig:goes_lc_deriv_wavelet} and \ref{fig:esp_lc_deriv_wavelet}.}}
    \label{fig:RHESSI_AIA_GOES_lc}
\end{figure*}

{The most likely interpretation of the second ridge in the wavelet power spectra in Figs.\ \ref{fig:goes_lc_deriv_wavelet} and \ref{fig:esp_lc_deriv_wavelet} is that some of the the peaks seen after $\sim$01:45 UT are from small flares in other active regions on the solar disk. They are revealed in the full-Sun images of SDO/AIA, GOES/SXI, and Hinode/XRT, and are listed in the Heliophysics Events Knowledgebase (HEK). \footnote{http://www.lmsal.com/hek/index.html}  For example, the GOES pulsations seen in Fig.\ \ref{fig:lc_hsi_goes_desmoothed}(b) between 01:46 and 01:52 UT are believed to result from a small flare in AR11745 at N13E24 seen in the Hinode/XRT image in Fig.\ \ref{fig:XRT_image} and in the light curve in Fig.\ \ref{fig:XRT_lc}. Two other flares from this active region are evident in the SDO/AIA images.  Their contributions to the AIA 131~\AA~flux from AR11745 are shown in Fig.\ \ref{fig:RHESSI_AIA_GOES_lc}.  As a consequence of the role of these small flares from other active regions, the second ridge in the wavelet power spectrum was not used in the analysis discussed below.}

\section{DISCUSSION}
\label{discussion}

The time derivatives of the {\em GOES} 1--8 and 0.5--4~\AA~light curves show significant structure extending from the start of the event at 00:00 UT for at least three hours well into the decay phase.  The wavelet analysis shows clear trends in the characteristic time scale of this structure that changes systematically from {$\sim$20~s at the time of the SXR peak at 01:10~UT to $\sim$100~s about an hour later (Figs.\ \ref{fig:goes_lc_deriv_wavelet} and \ref{fig:esp_lc_deriv_wavelet}). As indicated in Section \ref{Other ARs}, the second ridge in the wavelet power plots are the result of small flares from a different active region, and it is not included in the subsequent analysis.} The RHESSI 25--50~keV and higher energy X-ray light curves also show pulsations loosely correlated with the SXR structure similar to what would be expected from the Neupert Effect (Figs.\ \ref{fig:lc_hsi_goes}(a) and \ref{fig:lc_hsi_goes_desmoothed}(a)).  But what can account for the gradual trend to longer time scales during the extended decay of the event?  The HXR pulses and the associated pulses in the SXR time derivative are consistent with the concept of impulsive energy release that both accelerates electrons to tens of keV or higher to produce the HXRs and, perhaps simultaneously, heats plasma to $\ge$10 MK to produce the SXRs.  The accelerated electrons must heat plasma as they lose their energy by Coulomb collisions but whether this is the only source of plasma heating is not clear.  But what controls the time scale of the pulsations?  We explore the two possibilities mentioned in the introduction - (1) the time scale of the energy release process itself, and (2) the time scale of MHD waves in the magnetic loops formed after each energy release.

The time scale of the energy release process in the multi-island reconnecting system proposed by \cite{2006Natur.443..553D} depends on the Alfv\'{e}n speed in the vicinity of the reconnection site and the dimensions of the reconnection region. The Alfv\'{e}n speed is given by the following expression:
\begin{equation}
v_A = 2.18 \times10^{11}~(m_i/m_p)^{-1/2}~n_i^{-1/2}~B~cm~s^{-1}
\label{Eq:AlfvenSpeed}
\end{equation}
where $m_i$ is the ion mass, $m_p$ is the proton mass, $ n_i$ is the ion number density of the plasma in $cm^{-3}$, and B is the magnetic field strength in gauss. We use the mean coronal molecular weight, $m_i/m_p$, of 1.27 given in Appendix D of \cite{2005psci.book.....A}.

The density of the coronal SXR source can be estimated from the emission measure (EM) and the source volume ($V_m$).  The emission measure was obtained from the GOES data using the expressions given by \cite{2005SoPh..227..231W} that are incorporated in the SSW GOES Workbench.\footnote{http://hesperia.gsfc.nasa.gov/rhessidatacenter/complementary\_data/goes.html}  The volume was determined from RHESSI 6--12 keV images according to the following relation;
\begin{equation}
V_m = A~(\sigma_X + \sigma_Y)/2.0
\label{Eq:volume}
\end{equation}
where $A = \pi~\sigma_X \sigma_Y$ is the area of an ellipse with semi-axes computed from the second moments of the RHESSI image in the X-and Y directions, $\sigma_X^2$ and $\sigma_Y^2$.

The plasma density is then given be the following expression:
\begin{equation}
\rho=\sqrt{EM/(f~V_m)}
\label{Eq:density}
\end{equation}
where f is the filling factor relating the measured volume to the actual volume of the plasma.

\begin{figure}
    \includegraphics*[angle = 0, width = 0.5\textwidth]
    {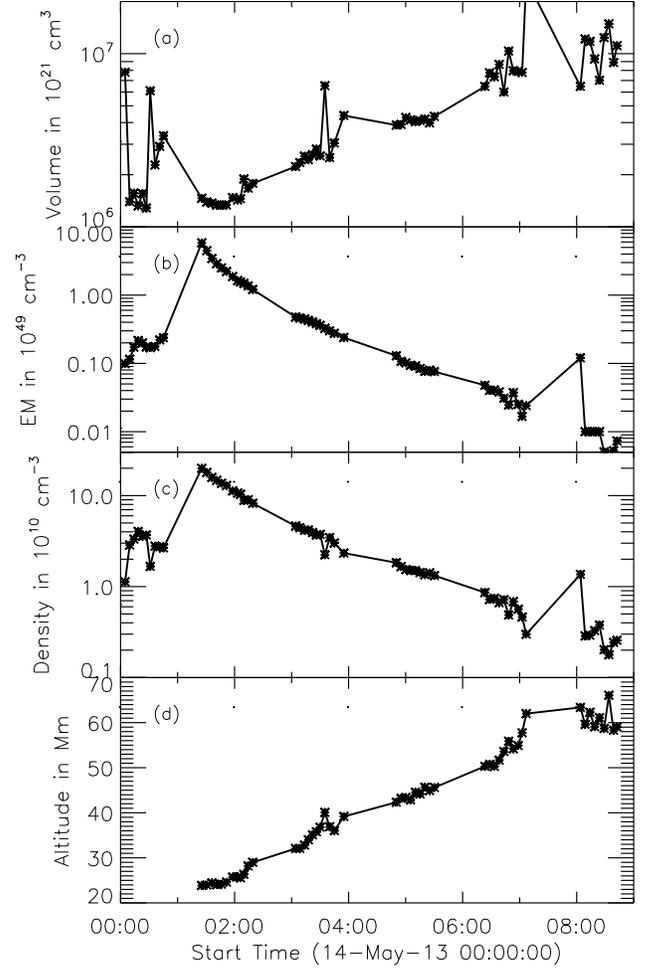}
    \caption{Time history of various parameters derived from the RHESSI and GOES observations. (a) Source volume determined from the RHESSI 6--12~keV images using Eq.\ \ref{Eq:volume}. (b) Emission measure derived from the GOES 2-channel SXR observations. (c) Source density derived from volume and emission measure shown in (a) and (b), respectively, using Eq.\ \ref{Eq:density} with unity filling factor. (d) Altitude estimated from the measured distance from the RHESSI 6--12~keV source centroid to a representative point halfway between the HXR foot points detected in Orbit~1.}
    \label{fig:params_vs_time}
\end{figure}

Values of the source volume, emission measure, and density (computed assuming unity filling factor) are plotted in Fig.\ \ref{fig:params_vs_time} for the full 8-hour duration of the event. At 02:18~UT, the time of the image in Fig. \ref{fig:im4_AIA_RHESSI}(c), the GOES emission measure was $1.5 \times 10^{49}~cm^{-3}$, the area of the RHESSI source was $2 \times 10^{18}~cm^2$, and the volume was $1.5 \times 10^{27}~cm^{3}$ giving a density of $\approx 10^{11}~cm^{-3}$.
No useful measures of the coronal magnetic field are available for this near-limb event but if we assume a value of B = 100 gauss at this time, then we obtain a reasonable Alfv\'{e}n speed of $700~km~s^{-1}$.

The relevant dimensions of the reconnection region or of the individual magnetic islands envisioned by \cite{2006Natur.443..553D} are difficult to estimate.  The distance traveled at the Alfv\'{e}n speed in $\sim$85~s, the characteristic time of the SXR structures at this time determined from the wavelet analysis (Fig. \ref{fig:goes_lc_deriv_wavelet}), is $700~km~s^{-1} \times 85 s = 6 \times 10^7~m = 60~Mm$. This is similar to the 44 Mm estimated length of the highest magnetic loops seen in AIA images at this time (see Section~\ref{Imaging}) but seems implausibly long for the dimensions of a magnetic island.  However, it is consistent with the alternative explanation for the trending time scales of the SXR structures that they are dependent on the MHD wave propagation in the magnetic loop(s) formed after each energy release occurrence.

The time to reach the footpoints from the 44 Mm high loop top at the Alfv\'en speed of $700~km~s^{-1}$ assuming a semicircular loop is\\
\indent$44 \times \pi/2 \times (10^8~cm) / (7\times10^7~cm~s^{-1})~=~\sim100~s$.\\
This is consistent with the time scale determined at this time during the flare from the wavelet analysis (Fig.\ \ref{fig:goes_lc_deriv_wavelet}) and the time between peaks (Fig.\ \ref{fig:time_between_GOES_peaks}).

The change in the time scale can also be compared for consistency with the change in height of the X-ray and EUV sources, or perhaps more realistically with the change in the length of the loops, as the flare progresses. From Fig. \ref{fig:hsi_source_location}, we see that the centroid of the RHESSI 6 - 12 keV source changed from an X--Y location of $[-940, 200]$ arcsec during Orbit 1 at 01:20 UT to $[-980, 235]$ arcsec after 08:00 UT in Orbit 5.  We can estimate what this means in terms of the change in the loop lengths by assuming that the footpoints of the loops were as initially located in the RHESSI 25 - 50 keV images during Orbit 1. After the end of the HXR emission before the start of Orbit 2, we assume that the loop footpoints remained on the ribbons seen at the AIA 1700~\AA~images.  Thus, initially, the altitude of the RHESSI source centroid was calculated from the measured distance between the location of the centroid and a point halfway between the footpoints, i.e.\ at [-905, 195] arcsec.  Thus, initially, the half-length of the loop was \\
\indent$\sqrt{(940-905)^2 + (200-195)^2} \times \pi/(2~cos10^{\circ}) \times 725~km = 41~Mm$\\
During RHESSI Orbit 5 after 08:00 UT, the loop footpoints had moved such that the point halfway between them was at $[-920, 190]$ based on the AIA 1700~\AA~images at that time.  Thus, the loop half-length is estimated to be \\
\indent$\sqrt{(980-920)^2 + (235-190)^2}  \times \pi/(2~cos10^{\circ}) \times 725~km = 87~Mm$

Following this general approach, knowing the plasma density and assuming that the relevant speed is the Alfv\'en speed given by the loop half-length divided by the time scale of the QPP, we can estimate the only unknown in Eq. \ref{Eq:AlfvenSpeed}, the magnetic field strength, B, as a function of time or altitude. We make the following assumptions:
\begin{enumerate}
\item The time scale, $\tau$ seconds, is given by the wavelet analysis.  Since the QPP are not detectable after 03:20~UT, we have linearly extrapolated the trend lines shown in Fig.~\ref{fig:goes_lc_deriv_wavelet} to 08:40~UT in RHESSI Orbit 5 in order to continue the estimate of B through the end of the event.
\item The length scale (L cm) is given as the loop half-length determined from the height (h) measured from the centroid location of the 6--12~keV X-ray source above the footpoints (Fig.\ \ref{fig:hsi_source_location}).  Assuming semicircular loops, $L=\pi~h/2$.
\item The ion density, $n_i$ is taken to be the density, $\rho$, determined from the GOES emission measure and the source volume $V_m$ obtained from RHESSI 6--12~keV images.
\item The relevant velocity ($v$) is the loop half-length divided by the time scale, and this is set equal to the Alfv\'en speed, i.e.

\begin{equation}
v = L/\tau = v_A = D~B~\rho^{-1/2}
\label{Eq:v}
\end{equation}
where $D = 1.93 \times 10^{11}$.

\end{enumerate}

With these assumptions we obtain the following expression for the magnetic field strength:
\begin{equation}
B = L~\rho^{1/2}/(D~\tau) = \pi~h~\rho^{1/2}/(2~D~\tau)
\label{Eq:B}
\end{equation}


If, instead of using the loop half-length as the relevant length scale, we assume that the QPP are the result of vertical kink-mode oscillations of the newly formed loops \citep[e.g.][]{1997SoPh..176..127M,2011ApJ...736..102A}, then we can relate the measured time scale $\tau$ to the period of these oscillations, $P_k$. In this case, the wavelength of the fundamental standing wave is twice the loop length (due to the forward and backward propagation). When the the plasma density outside the loop is negligible compared to the density inside the loop (as expected for this flare), $P_{k} = 4 L/(\sqrt{2} v_{A}$) \citep[Eq. 7.2.4 of][]{2005psci.book.....A}. Thus, setting $\tau = P_k$ and substituting for L and $v_A$, we get the following expression for the magnetic field strength as a function of the measured source altitude (h), density ($\rho$), and time scale ($\tau$):
\begin{equation}
B = \sqrt{2}~\pi~h~\rho^{1/2}/(D~\tau)
\label{Eq:B_kink}
\end{equation}

\begin{figure}
    \includegraphics*[angle = 0, width = 0.5\textwidth, trim = 0 50 0 50]
	{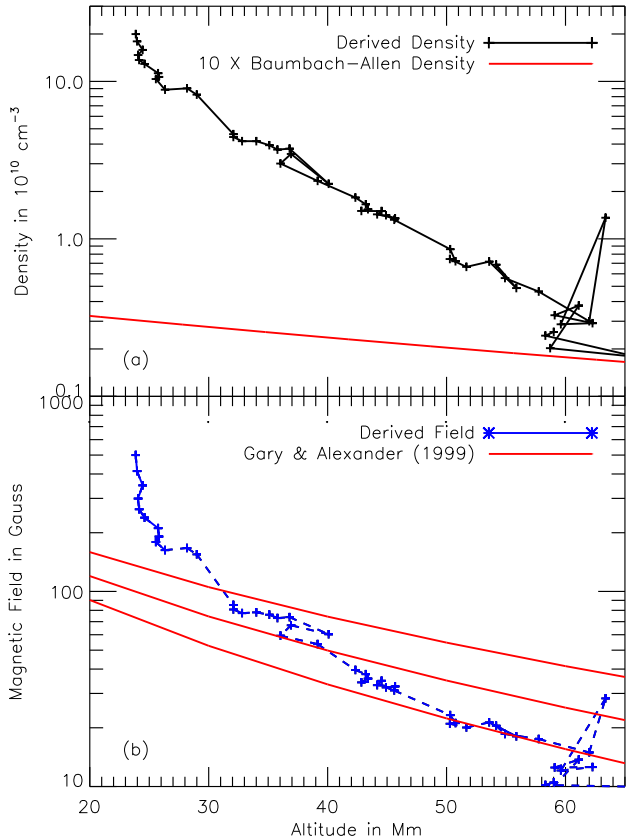}
    \caption{Derived density and magnetic field strength plotted as a function of the estimated altitude of the RHESSI 6--12~keV source. (a) The same plasma density plotted in Fig. \ref{fig:params_vs_time}(c).  Also shown for comparison is ten times the density given by the Baumbach--Allen formula \citep{2000asqu.book.....C}, often taken to represent the expected values above an active region. (b) Magnetic field strength estimated using Eq.\ \ref{Eq:B_kink}. {The value at each altitude was calculated using the first wavelet trend line shown in Figs. \ref{fig:goes_lc_deriv_wavelet} and \ref{fig:esp_lc_deriv_wavelet}. This line was linearly extrapolated to longer time scales after the pulsations were no longer detected, i.e. for altitudes above $\sim$26~Mm.} Also shown (in red) are the values computed from Eq.\ \ref{Eq:B_G&A} for $\zeta$ = 1.3 (top), 1.5 (middle), and 1.7 (bottom).
 }
    \label{fig:nB_vs_alt}
\end{figure}

The estimated magnetic field strength computed in this way is plotted vs.\ source altitude in Fig.\ \ref{fig:nB_vs_alt}(b).


The calculated densities shown in Fig.\ \ref{fig:params_vs_time}(c) and \ref{fig:nB_vs_alt}(a) range from $2 \times 10^{11}$~to~$2 \times 10^9~cm^{-3}$~and are consistent with previous estimates for other events. They are initially almost two orders of magnitude higher than the 10 x Baumbach-Allen values but decrease to almost the same level towards the end of the flare.  These higher densities presumably reflect the compression produced by the magnetic reconnection and the back filling by chromospheric evaporation.  The magnetic field strengths shown in Fig.~\ref{fig:nB_vs_alt}(b) range from {500} gauss at 24 Mm to {10--20} gauss at 60 Mm.  They are also consistent with previous estimates and show a similar trend with altitude (h) as the empirical function
\begin{equation}
    B = 1000~\exp(-\zeta~\sqrt{h})
    \label{Eq:B_G&A}
\end{equation}
given by \cite{1999SoPh..186..123G}, where h is in units of $10^4~km$ and $\zeta$ is between 1.3 and 1.7.


One difficulty with the MHD wave explanation is the small amplitude and the expected constancy of the oscillations in the spatially integrated SXR intensity that results from any of the possible wave modes. It was originally thought that kink mode waves do not result in any change in the total X-ray emission from the plasma in the loop \citep[Section 7.3.3]{2005psci.book.....A}.  This is correct to first order for horizontal kink modes, where the loop just moves horizontally from side to side with no change in the volume or density of the plasma in the loop. However, in the vertical kink mode, where the loop top moves up and down, the loop length changes. In addition, the loop transverse diameter also changes because of the conservation of magnetic flux.  Thus, the volume, and hence the electron and ion densities, change so that the emission measure also changes since it depends on the integral of the density squared over the volume. This gives a commensurate change in the X-ray flux \citep{2011ApJ...736..102A}. The changes are expected to be a small fraction of the steady emission, and this is observed.

The observed small amplitude must partly be because of the steady emission from previously formed loops that are still hot enough to emit in the GOES wavebands. The expected constancy of the emission is also a problem since observationally the pulses seem to come and go, especially later in the decay phase after {02:00~UT}, when they cease completely or their amplitude becomes so low that they are undetectable.  The steady emission continues for another {6} hours and the source continues to rise to higher altitudes suggesting continuous reconnection and the formation of new hot loops.

One possible explanation for the absence of pulsations after {02:00~UT} is that the fast sausage mode is a contributor in addition to or instead of the vertical kink mode. This idea is supported by the recent report of loop oscillations detected with the Interface Region Imaging Spectrograph \citep[IRIS,][] {2014SoPh..289.2733D} that have been attributed to sausage mode oscillations \citep{2016ApJ...823L..16T}. The case for the sausage mode rather than the kink mode is strengthened by {their} measurement of a $\pi/2$ phase shift between the Doppler shift oscillation and the GOES intensity oscillations. 

The analysis of sausage mode oscillations is less straightforward than for kink mode oscillations.  This is because the characteristic time scale of the sausage mode also depends on the Alfv\'en speed external to the loop, which is likely to be much higher than the internal Alfv\'en speed because of the lower density. This could provide an explanation for the shorter time scales of the QPP early in the event. A further complication is that standing sausage modes are separated into trapped and leaky categories depending on the ratio of loop length (L) to transverse radius (a). Thus, for a given radius, the period of the trapped sausage mode waves increases with increasing loop length until the critical L/a ratio is reached and the mode becomes leaky \citep[e.g.][]{2012A&A...543A..12G, 2014A&A...568A..31L}.  This could explain the increasing {time scales} measured early during the decay of the flare and the disappearance of the pulsations {from AR11748 after 02:00~UT}. The altitude of the X-ray source at that time was {$\sim$26~Mm} corresponding to a loop length of {$L = \pi \times h = 80$~Mm}.  For a typical transverse loop radius of 1 Mm, this would give a cutoff ratio {$L/a \approx 80:1$}.  This value is outside the $\sim6 - 20$ range of allowed values given in Section 7.3.1 of \cite{2005psci.book.....A} for ratios of the internal ($n_0$) to external ($n_e$) loop densities of $n_0/n_e \approx10^2~-~10^3$.  Thus, this suggests that the sausage mode is not the dominant cause of the observed pulsations.

A possible model to explain how MHD waves are initiated involves the variability of both the magnetic field and the density of the plasma flowing into the current sheet. \cite{2012ApJ...745L...6T} reported that the inflow speed changed from 90 to 12~$km~s^{-1}$ in 5 minutes during a flare. \cite{2013NatPh...9..489S} obtained inflow speeds of 20--70~$km~s^{-1}$ for a different flare.  These speeds are much lower than the Alfv\'en speed expected at an altitude of $\sim$20~Mm, where these observations were made. The 10-times Baumbach-Allen density above the active region is $\sim3~10^9~cm^{-3}$ and the magnetic field strength from \cite{1999SoPh..186..123G} is $\sim$100~G, giving an Alfv\'en speed of $\sim4000~km~s^{-1}$.  The inflow speed is presumably controlled by the difference between the pressure in the current sheet and the pressure in the inflow region. It is this differential pressure that sustains the reconnection and determines the reconnection rate as the flare progresses to higher altitudes.  While any changes in the density or magnetic field of the inflow will affect the energy release rate, it is difficult to see how it would produce QPP, particularly with time scales that depend on altitude as observed.  Thus, it is probable that these effects produce bursts of increased energy release that could trigger kink mode and/or sausage mode oscillations of the newly formed loops, as suggested by the comparison of HXR and SXR light curves in Figs.\ \ref{fig:lc_hsi_goes} and \ref{fig:lc_hsi_goes_desmoothed}, but it is unlikely that they could be solely responsible for producing the observed SXR QPP.

Positively identifying the underlying production mechanism will be required before the full diagnostic potential of QPP can be realized.  It is not clear if that can be achieved with the current observations although the measurement of Doppler shift oscillations with IRIS \citep{2016ApJ...823L..16T} certainly helps. At present, only constraints can be applied based on evaluating various assumed oscillating modes  and comparing predicted QPP properties with observations.

\section{SUMMARY AND CONCLUSIONS}
\label{conclusions}

We have detected and identified 163 pulses in the time derivative of the GOES light curves during the decay of the X3.2 flare on 2013 May 14. Near identical peaks are evident in the time derivative of the EVE/ESP light curve showing that these features must be of solar origin and not some instrumental artifact. Most of the pulses occurred in the first two hours during and after the HXR impulsive phase of this event, which lasted for a further 5 hours before the SXR flux returned to preflare levels.  Using both a wavelet analysis technique and simply locating the peaks by eye, we have shown that the time scale or time between peaks increases with time along two trend lines.  The time scale along the first line changes from $\sim$10~s during the impulsive phase starting at 01:00 UT to $\sim$100~s at 02:00~UT; the time scale along the second trend line changes from $\sim$40~s at 01:40~UT to $\sim$90~s at 03:00~UT.  {This second trend line results from small flares in a differnet active region and is not used in the subsequent analysis.}

Although RHESSI was at nighttime during the SXR rise and peak of this event, RHESSI images provide the locations and extent of both HXR and SXR sources at later times. Two HXR (25--50~keV) footpoint sources were detected during the first orbit when RHESSI observed the event between 01:24 and 02:22~UT. SXR (6--12~keV) sources were imaged during the daytime parts of five RHESSI orbits starting at 01:20~UT, after the GOES peak at 01:11~UT, and extending until 08:40~UT (Fig.~\ref{fig:im4_AIA_RHESSI}).  During this time the position of the source centroid moved along a near linear path with a position angle of $48.5^\circ$ from north at a speed of $1.7~km~s^{-1}$ (Fig.~\ref{fig:hsi_source_location}). The source volume estimated from the images increased from $1.3~10^{27}~cm^{3}$ at 01:20 UT to $10^{28}~cm^{3}$ at 08:40 UT (Fig.~\ref{fig:params_vs_time}(a)). Combined with the emission measure determined from the GOES data (Fig.~\ref{fig:params_vs_time}(b)), this volume allows the plasma density to be determined assuming unity filling factor. It was found to decrease from $2~10^{11}~cm^{-3}$ at 01:20 UT to $\sim2~10^9~cm^{-3}$ at 08:00 UT (Fig.~\ref{fig:params_vs_time}(c). Since the event was close to the solar limb, an estimate could be made of the source altitude above the footpoints (Fig.~\ref{fig:params_vs_time}(d)) that ranged from $\sim$24~Mm at 01:20~UT to $\sim$60~Mm at 08:00 UT. The calculated density falls exponentially with the altitude determined in this way as expected (Fig.~\ref{fig:nB_vs_alt}) but the e-folding scale length is significantly smaller (21 Mm) than the value for the Baumbach-Allen model (71 Mm) over the covered altitude range.  The higher density above the ambient value presumably results from the magnetic reconnection and plasma heating during the flare. The more rapid decrease in density with altitude may reflect the decrease in the energy released as the reconnection site progresses to higher altitudes and weaker magnetic fields.

With the assumption that the observed QPP were produced by vertical kink mode oscillations in newly formed magnetic loops, we are able to calculate the coronal magnetic field as a function of altitude.  By setting the expected period of the kink modes equal to the measured time scale of the QPP taken from Fig.~\ref{fig:goes_lc_deriv_wavelet}, we obtained Equation \ref{Eq:B_kink}. This defines the magnetic field strength in terms of measured quantities - the source density and altitude, and the time scale. The plasma density in the SXR source is determined from the emission measure derived from the GOES two-channel observations and the source volume from the RHESSI 6--12~keV images. This density is plotted in Fig.\ \ref{fig:nB_vs_alt}(a) as a function of the source altitude, also determined from RHESSI images.  The resulting magnetic field strength is plotted in Fig.~\ref{fig:nB_vs_alt}(b) as a function of altitude. The values were extended beyond the time when QPP were visible in the GOES light curves by linearly extrapolating the measured variations of QPP time scales to the end of the event.  The magnetic field strength calculated in this way is 200--500~G at 24 Mm decreasing to {10--20~G} at 60 Mm. 
Also plotted in Fig.~\ref{fig:nB_vs_alt}(b) is the coronal magnetic field strength computed by \cite{1999SoPh..186..123G} for a simple bipolar active region using photospheric magnetograms and observed coronal loops.  Applying this method in our case would be difficult since the active region is within $10^\circ$ of the solar east limb.

The consistency of the estimated magnetic field strength with expectation values during the decay phase of this event shows the viability of the assumption that the QPP are produced by vertical kink mode oscillations of newly formed loops. If the relation between the timescales of the SXR variability can be more firmly tied to the dimensions of the emitting plasma and hence to the local Alfv\'en speed, then this method may become a useful way of determining the magnetic field strength in the corona at or near the site of continued energy release. For the present, the general reasonableness of the calculated densities and magnetic field strengths and their variations with altitude show that the kink mode interpretation of the SXR QPP in this event is tenable. However, the sausage mode is not ruled out as the source of the measured oscillations. It needs to be further explored, especially in light of the recent IRIS  observations reported by \cite{2016ApJ...823L..16T} that strongly support the sausage mode interpretation. 

The size scale of the energy release site is likely to be much shorter than the loop length leading to unreasonably small magnetic field strengths. Hence, it seems less likely that multiple bursts of energy release are the direct cause of SXR QPP, at least during the decay phase of the event.  During the impulsive phase, however, the amplitude of any effects from kink or sausage mode oscillations may be swamped by much greater amplitude variations in the energy release rate itself.

The type of wavelet analysis used here will be applied to many other GOES events to determine if the trends to longer time scales are present in all cases or only in those larger events with long decay times. Clearly, a more realistic model of the processes that could produce the observed QPP should be investigated to the extent that is warranted by the quality of the observations.  The density of cooler plasma in the loops must be included based on the differential emission measure that can be determined from EUV imaging observations from AIA and other instruments.  The distribution of plasma densities both along the bright loops and external to the loops containing the hot plasma must also be considered in estimating expected QPP characteristics. Results from QPP analysis for events nearer to disk center should also be compared with magnetic field extrapolations from photospheric magnetograms. Thus, consistency of the observations with a more detailed model must be further investigated before the accuracy and reliability of the derived coronal magnetic field strengths can be evaluated.  However, the initial values obtained here using the simplest MHD model show the viability of the method and its potential usefulness in providing unique diagnostics of the flaring plasma and the environment in which it is produced.

Further advances in our understanding of these structures in the SXR light curves will require new high time resolution imaging and spectroscopy observations to determine the location and evolution of each SXR structure and its relation to any corresponding HXR pulses.  The Focusing Optics X-ray Solar Imager \citep[FOXSI,][]{2013SPIE.8862E..0RK, 2014ApJ...793L..32K} has the required capabilities, and is being planned for observations during the next solar maximum.  It will make images with $\sim$5 arcsec angular resolution, 0.5 keV energy resolution, and sub-second time resolution in the energy range from $\sim$3 to $\gtrsim$60~keV.  Thus, QPP in events of the type presented in this paper could be imaged at different energies to determine their location within the flaring region, and followed on time scales shorter than their apparent evolution time scales, even during the impulsive phase.  This should allow a more definitive determination to be made of the origin of QPP, either as the direct result of the time scale of the energy release process itself or of the MHD wave propagation in magnetic loops formed following magnetic reconnection.

\acknowledgments

The work of T.W. was supported by the NASA Cooperative Agreement NNG11PL10A to the Catholic University of America.

{\sc chianti} is a collaborative project involving the US Naval Research Laboratory, the Universities of Florence (Italy) and Cambridge (UK), and George Mason University (USA).

IDL wavelet software was provided by C. Torrence and G. Compo,
and is available at URL: http://paos.colorado.edu/research/wavelets/.

 The SOHO LASCO CME catalog is generated and maintained at the CDAW Data Center by NASA and The Catholic University of America in cooperation with the Naval Research Laboratory. SOHO is a project of international cooperation between ESA and NASA.


{\em Facilities: RHESSI, GOES}

\appendix

\section{Wavelet Analysis}
\label{wavelet}

We have used a modified version of the IDL wavelet software discussed by \cite{1998BAMS...79...61T}.  The package computes the wavelet power equal to the square of the convolution of the wavelet function with the data for the range of periods allowed by the data.  We used the Morlet wavelet, a plane wave modulated by a Gaussian with the following form:
%
\begin{equation}
\Psi_0(\eta) = \pi^{-1/4}~e^{6~i~\eta}~e^{-\eta^2/2}
\end{equation}
Here, $\Psi_0$ is the amplitude of the wavelet and $\eta$ is the the parameter over which the data varies (time in our case) in units of the standard deviation $\sigma$ of the Gaussian.

The wavelet transform of the discrete sequence of data values $x_n$ is defined as the convolution of $x_n$ with a scaled and translated version of the wavelet as follows:

\begin{equation}
W_n(s) = \sum_{n'=0}^{N-1} x_{n'} \Psi^{ *}\left[\frac{(n' -n)\delta t}{s}\right]
\end{equation}

where the (*) indicates the complex conjugate and $s$ is the wavelet scale. By
varying $s$ and translating along the localized time index, $n$, an array can be obtained of the complex wavelet transform, W = $\Re(W) + \Im(W)~i$, as a function of both s and time. The wavelet transform amplitude is $\Re(W)$, the phase is $\arctan(\Im(W), \Re(W))$, and
the wavelet power spectrum is $|W_n(s)|^2$.

 The integral of this wavelet over all times is identically zero, Thus, if the data are constant within a time range extending for  $\sim3\sigma$ on either side of the Gaussian peak time, the wavelet transform amplitude will be essentially zero. But any variation in the signal over that time range will result in a positive or negative value of $W_n(s)$ and hence a positive power.  The wavelet power is computed for all times for which there is data and for a range of values of $\sigma$ allowed by the data.  The wavelet timescale, s, is taken to be $6\sigma$ and ranges from half the duration of the data to twice the cadence. Edge effects cause the power to be unreliable for one period after the start and one period before the end of the data set. This range of exclusion is very small in our case for the periods of interest that are much smaller than the over 2-hour duration of the observations. It is the region outside of the two sloping white lines close to the left and right edges of the wavelet power plot in the bottom of Fig.~\ref{fig:goes_lc_deriv_wavelet}.

 It is known that there is a bias in the wavelet power computed using the \cite{1998BAMS...79...61T} software - two signals with the same amplitude but different frequencies produce different wavelet power. This arises because of the difference between wavelet analysis and Fourier analysis.  In Fourier analysis the power for each value of the period (or frequency) is accumulated over the full duration of the data set.  In wavelet analysis, by contrast, the power is accumulated only over the effective duration of the wavelet, which for a Morlet wavelet is essentially $6\sigma$ and so is different for each timescale.
 \cite{2007JAtOT..24.2093L} have shown that this bias can be readily overcome by dividing the wavelet power by the timescale used to compute it.  In this way, the power per unit time is obtained for each observing time interval as opposed to the power averaged over the full duration of the data set.

 A second difference between wavelet and Fourier analysis is their response to incoherent signals at a given frequency where the phase of the signal varies with time.  With wavelet analysis, the power summed over time is independent of any phase change in the signal whereas with Fourier analysis the computed power is decreased by the changes in phase of the signal. Thus, the summed wavelet power will be greater than the Fourier power depending on the level of incoherence in the signal.  This is particularly important in the analysis of solar flare data where the emissions at different times can come from different locations that are not expected to maintain any phase relation with one another. Thus, it is not expected that phase coherence will be maintained for times longer than the duration of elementary flare bursts that typically last for seconds to tens of seconds.

 A third issue with the standard wavelet analysis is that the power, or power per unit time, is dependent on the amplitude of variations in the data.  This can make it difficult to see the presence of significant QPP if the amplitude of the variations of interest changes during the duration of the data set.  This problem can be overcome to a large extent by dividing by the smoothed data such that the signal of interest has a largely constant amplitude for the duration of the data.

\bibliographystyle{apj}	

\bibliography{qpp_21march2016}

\end{document}